\documentclass[sn-mathphys-num]{sn-jnl}

\usepackage{graphicx}%
\usepackage{multirow}%
\usepackage{amsmath,amssymb,amsfonts}%
\usepackage{amsthm}%
\usepackage[title]{appendix}%
\usepackage{xcolor}%
\usepackage{algorithm}%
\usepackage{algorithmicx}%
\usepackage{algpseudocode}%
\usepackage{subcaption}
\usepackage{cleveref}

\usepackage{hhline}
\usepackage{afterpage}
\usepackage{pdflscape}
\usepackage{orcidlink}

\newtheorem{definition}{Definition}%

\begin{document}

\title[Fast maximal clique enumeration in weighted temporal networks]{Fast maximal clique enumeration in weighted temporal networks}

\author*[1]{\fnm{Hanjo D.} \sur{Boekhout}\orcidlink{0000-0002-8456-9063}}\email{h.d.boekhout@liacs.leidenuniv.nl}

\author[1]{\fnm{Frank W.} \sur{Takes}\orcidlink{0000-0001-5468-1030}}\email{f.w.takes@liacs.leidenuniv.nl} 

\affil[1]{\orgdiv{LIACS}, \orgname{Leiden University}, \orgaddress{\street{Einsteinweg 55}, \postcode{2333 CC}, \city{Leiden}, \country{The Netherlands}}}

\abstract{
Cliques, groups of fully connected nodes in a network, are often used to study group dynamics of complex systems. 
In real-world settings, group dynamics often have a temporal component. For example, conference attendees moving from one group conversation to another.
Recently, maximal clique enumeration methods have been introduced that add temporal (and frequency) constraints, to account for such phenomena. These methods enumerate so called $(\delta,\gamma)$-maximal cliques.

In this work, we introduce an efficient $(\delta,\gamma)$-maximal clique enumeration algorithm, that extends $\gamma$ from a frequency constraint to a more versatile weighting constraint. 
Additionally, we introduce a definition of $(\delta,\gamma)$-cliques, that resolves a problem of existing definitions in the temporal domain.
Our approach, which was inspired by a state-of-the-art two-phase approach, introduces a more efficient initial (stretching) phase. Specifically, we reduce the time complexity of this phase to be linear with respect to the number of temporal edges.
Furthermore, we introduce a new approach to the second (bulking) phase, which allows us to efficiently prune search tree branches. Consequently, in experiments we observe speed-ups, often by several order of magnitude, on various (large) real-world datasets.
Our algorithm vastly outperforms the existing state-of-the-art methods for temporal networks, while also extending applicability to weighted networks.}


\maketitle
\section{Introduction}
\label{sect:intro}

To study real-world systems, the field of network science models these systems as networks of connected entities.
For example, entities, also called nodes, can represent the attendees at conferences that are connected through their proximity~\cite{Isella:2011qo}, online social media users connected through their interactions~\cite{panzarasa2009patterns,viswanath2009evolution}, or online communities connected through hyperlinks~\cite{kumar2018community}. 
Finding groups of entities that are fully connected to one another, called \emph{cliques}, in these networks, can provide valuable insight into (the group-dynamics of) the complex systems they model. 
As such, the enumeration of cliques has long been the subject of study in network science and graph theory~\cite{leon2019enumerating}.

Many real-world systems are more complex than can be modeled by a simple (undirected) graph. 
Connections may be short-lived, occurring only at specific times, such as the proximity of conference attendees.
Different connections may also hold differing significance. For example, the hyperlinks between online communities may be accompanied by positive or negative sentiment. 
We can model such complexities through timestamps and weights on the connections to form, so called, \emph{(weighted) temporal networks}.
Consequently, we can also add temporal and weighting constraints on the clique enumeration.
In other words, we may require cliques to always be fully connected with a given minimum weight $\gamma$ for every $\delta$-time interval of their time span.
In the example of the social network modeling people in proximity of each other, we would be looking at people that interacted with each other a particular number of times $\gamma$ during each time window $\delta$ in a particular time span. 
Specifically, this work focuses on the enumeration of \emph{$(\delta,\gamma)$-maximal cliques}, which are $(\delta,\gamma)$-cliques for which neither the node set nor the time span of the clique can be extended further.

In the literature, several methods have been proposed to enumerate temporal cliques (i.e., $\delta$-cliques).
In a recent publication, Banerjee \& Pal~\cite{banerjee2024two} introduced a two-phase approach to enumerate $(\delta,\gamma)$-maximal cliques, where $\gamma$ referred to the frequency of links instead. 
Of course, for unweighted networks, when all edges have a weight of one, the frequency and cumulative weight are equal.
Here, we define $(\delta,\gamma)$-cliques to cover weighted networks, while fixing a practical problem of existing definitions in the temporal domain.
Additionally, we propose a different usage of $\delta$ in the definition of cliques to more naturally align with common-sense interpretation, to prevent incorrect usage by end-users.
Furthermore, we introduce a new $(\delta,\gamma)$-maximal clique enumeration algorithm that was inspired by Banerjee \& Pal's two-phase approach, but uses new approaches for both phases.
Importantly, our algorithm improves the time complexity of the first phase (i.e., the stretching phase), such that it is independent of $\gamma$, and introduces pruning of search tree branches to speed up the second phase (i.e., the bulking phase).

Through experiments on a set of both small and larger real-world network datasets with up to 4.4 million nodes and 19 million edges, we show that our proposed algorithm outperforms all existing temporal clique enumeration algorithms, often by several orders of magnitude, demonstrating better scalability.
In almost all cases, we outperform the current state-of-the-art approach by Banerjee \& Pal in both time and space complexity, and consequently, running time and memory usage.
 
The remainder of this paper is structured as follows. 
In section~\ref{sect:relwork} we discuss related work on clique enumeration.
Then, in section~\ref{sect:backg} we provide notation and definitions used throughout the paper, including our proposed definition of $(\delta,\gamma)$-cliques.
Next, we introduce our new faster $(\delta,\gamma)$-clique enumeration algorithm in section~\ref{sect:methods}.
In section~\ref{sect:data}, we describe the datasets used in our experiments.
Subsequently, the experiments and results are discussed in section~\ref{sect:experiments}.
Finally, we draw our conclusions in section~\ref{sect:conclusions}.

\section{Related Work}
\label{sect:relwork}
Cliques and their enumeration were already subject of research over 50 years ago. 
In 1973, Bron \& Kerbosch~\cite{bron1973algorithm} presented two backtracking algorithms for clique enumeration in simple static undirected networks, that use a branch-and-bound technique to prune search tree branches.
These algorithms still form the basis of some state-of-the-art approaches.
Tomita et al.~\cite{tomita2006worst} vastly improved the pruning of Bron \& Kerbosch's backtracking algorithms through smart pivot selection.
When expanding a clique, they prioritize vertices that have the largest neighborhood overlap with the set of candidate vertices. 
Consequently, Eppstein et al.~\cite{eppstein2010listing} employed a degeneracy ordering in the outer level of recursion of Tomita's adaptation, to achieve a time complexity that is nearly worst-case optimal.
These adaptations of the Bron \& Kerbosch algorithm, are still used in many network science software packages today.
The pruning techniques we apply in this paper for temporal cliques, were in part inspired by these methods.

Various other methods and problem extensions were introduced over the years. 
For example, some approaches explored the use of partitioning to deal with limited memory~\cite{cheng2012fast}, others attempted to improve efficiency through parallelisation~\cite{schmidt2009scalable,wu2009distributed,hou2016efficient,das2018shared}, and yet others tried using iterative enumeration~\cite{kose2001visualizing} or decomposition~\cite{manoussakis2019new}.
Furthermore, there were also works that, instead of maximal cliques, specifically considered maximum cliques~\cite{eblen2012maximum} or the slightly different isolated, pseudo, and/or defective cliques~\cite{ito2009enumeration,uno2010efficient,dai2023maximal} or even $k$-plexes~\cite{conte2017fast}.
Finally, some research focused on different network types, ranging from spatial networks~\cite{zhang2019efficient}, to uncertain networks~\cite{mukherjee2016enumeration}, to signed networks~\cite{chen2020efficient}, to temporal networks~\cite{viard2015revealing, viard2016computing,himmel2017adapting,viard2018enumerating,zhu2019scalable,qin2019mining,banerjee2019enumeration,banerjee2021two,banerjee2022efficient,banerjee2024two}.
In this work, we focus specifically on the latter type of network, the temporal networks and specifically $(\delta,\gamma)$-maximal cliques.

The $\delta$-maximal cliques, which enforce temporal constraints upon the edges of a clique, were first introduced by Viard et al.~\cite{viard2015revealing, viard2016computing}.
The methods introduced by Viard et al. to enumerate these cliques were greedy in nature and scaled poorly.
Himmel et al.~\cite{himmel2017adapting} then adapted the Bron \& Kerbosch algorithm to account for the temporal dimension to enumerate $\delta$-maximal cliques, which significantly improved on Viard's methods. 
Next, Banerjee \& Pal~\cite{banerjee2019enumeration} extended the $\delta$-maximal clique definition with a minimum edge frequency $\gamma$ requirement.
Recently, the same authors introduced a two-phase approach to more efficiently enumerate such ($\delta,\gamma$)-maximal cliques, which also improved upon Himmel et al.'s performance for $\delta$-maximal cliques~\cite{banerjee2021two,banerjee2024two}.
In their approach, the second phase relies on iterative enumeration which prunes only duplicate branches.
Here, we build upon the work of Banerjee \& Pal~\cite{banerjee2024two} to accommodate also weighted temporal networks and to improve the algorithm's efficiency through further pruning methods.
\section{Preliminaries} 
\label{sect:backg}
In this section we first introduce notation and definitions dealing with (weighted) temporal networks in section~\ref{subsect:backg-notation}. 
Next, in section~\ref{subsect:backg-max} we discuss existing maximal clique definitions (both static and temporal) and identify two practical problems existing definitions cause in the temporal domain when applied in a real-world setting.
Additionally, we redefine maximal cliques in the temporal domain such that these issues no longer occur and such that it can be applied to weighted temporal networks.
\subsection{Basic notation and definitions}
\label{subsect:backg-notation}
A \emph{temporal network} $\mathcal{G} = (V, E, \mathcal{T})$, is comprised of its static elements $V$ and $E$, describing the node set and the edges connecting those nodes respectively, and a mapping $\mathcal{T}$ of the edges to their temporal occurrence time stamps, i.e., $\mathcal{T}: E \mapsto 2^T\setminus\emptyset$ with $T = \{t_{min}, \cdots, t_{max}\}$ as the set of all discrete time stamps.
As such, the set of temporal edge occurrences $E^{\mathcal{T}}$ consists of all \emph{temporal edge instances} $(t,u,v)$, where $u,v \in V$, $(u,v) \in E$, and $t \in T$. 
We say that $m = |E^{\mathcal{T}}|$.
Note that $(u,v)$ does not imply a direction here, as we only consider undirected networks in this work.

Given the above, we can define a \emph{weighted temporal network} $\mathcal{G} = (V, E, \mathcal{T}, \mathcal{W^T})$, as a temporal network with the added mapping $\mathcal{W^T}: E^{\mathcal{T}} \mapsto \mathbb{R}$, which maps each temporal edge occurrence to a weight. 
The set of weighted temporal edge occurrences $E^{\mathcal{TW^T}}$, thus consists of all \emph{weighted temporal edge instances} $(t,u,v,w)$, where $u,v \in V$, $(u,v) \in E$, $t \in T$, and $w \in \mathbb{R}$. Note that a temporal network is equal to a weighted temporal network with all weights set to one.

Given an edge $(u,v)\in E$ and time interval $[t_{b}, t_{e}]$, the \emph{frequency} $f((u,v), [t_{b},t_{e}])$ represents the cumulative weight of all weighted temporal instances of edge $(u,v)$ in the interval $[t_{b}, t_{e}]$. 
As such, for interval [$t_{min}$, $t_{max}$], i.e., for the entire \emph{lifetime of the network}, the frequency of an edge in an unweighted temporal network equals its number of temporal occurrences.

We visually represent (weighted) temporal networks using the link stream model which shows the (weighted) relationship between nodes over time (see Figure~\ref{fig:failure}a).
\subsection{Maximal cliques}
\label{subsect:backg-max}
In a static network, a set of nodes $X \subseteq V$ is a \emph{clique} when they are fully connected, i.e., for all ${u,v \in X}$ it holds that $(u,v)\in E$.
We say that a clique is \emph{maximal}, when there exists no set of nodes $Y \subseteq V$ such that $X \subset Y$ and $Y$ is a clique.
This notion of (maximal) cliques was first extended by Viard et al.~\cite{viard2015revealing, viard2016computing} to the temporal domain as follows. \vspace{5pt}

\begin{definition}[$\delta$-clique]\label{def:d-clique}
    Given a time duration $\delta \in \mathbb{Z}+$, a $\delta$-clique of the temporal network $\mathcal{G}$ is a vertex set and time interval pair $C = (X, [t_b, t_e])$ with $X \subseteq V_{\mathcal{G}}$ and $t_b,t_e \in T$ such that for all $u,v \in X$, where $v \neq w$, and $\tau \in [t_b, t_e-\delta+1]$, it holds that $f((u,v), [\tau, \tau + \delta - 1]) \geq 1$. 
\end{definition} \vspace{5pt}

\begin{definition}[$\delta$-maximal clique]\label{def:dmax-clique}
    A $\delta$-clique $C = (X, [t_b, t_e])$ is considered \emph{$\delta$-maximal}, when there exists no $\delta$-clique $C' = (X', [t'_b, t'_e])$ such that either $X \subset X'$, $t'_b \leq t_b$, and $t'_e \geq t_e$ holds or $X = X'$ and $t'_b < t_b$ or $t'_e > t_e$ holds.
\end{definition} \vspace{5pt}

In other words, the vertex set $X$ of a $\delta$-clique is fully connected during every $\delta$-time interval in the interval $[t_b, t_e]$; and a $\delta$-maximal clique can neither extend its time interval for the current node set nor extend the node set without shrinking the interval.
In line with Banerjee \& Pal~\cite{banerjee2024two}, we refer to cliques that cannot extend their time interval, regardless of whether the node set is maximal, as \emph{duration-wise maximal}.

Note that, compared to previous related work, we changed the use of $\delta$ in the definition of a $\delta$-clique, to more closely align with its (natural) interpretation.
The intended interpretation of $\delta$, as also stated textually by previous works, is that ``every node pair must be connected at least after every $\delta$ timestamps in the time interval''.
Consider a periodic network with measurements, i.e., edges, occurring every 20 seconds.
Given such a network, the periodicity of found cliques has three phases: (1) \emph{pre-periodicity}, where $\delta$ is too small to connect subsequent events at timestamps 20 and 40 within one clique; (2) \emph{exact periodicity}, where for every edge in the clique both timestamps 20 and 40 are required to establish a clique over time interval [20, 40]; and (3) \emph{overlap periodicity}, where some edges having an occurrence at timestamp 20 and others at 40 is sufficient for a clique with time interval [20, 40].
Given the interpretation of $\delta$ above, the natural assumption would be that, $\delta \leq 19$ corresponds to pre-periodicity, $\delta = 20$ to exact periodicity, and $\delta \geq 21$ corresponds to overlap periodicity.
Although this holds for Definition~\ref{def:d-clique}, this did not hold for previous works where $\delta = 20$ corresponded to overlap periodicity and $\delta = 19$ to exact periodicity~\cite{himmel2017adapting,banerjee2019enumeration,banerjee2021two,banerjee2022efficient,banerjee2024two}.

\subsubsection{Solving a practical problem in the temporal domain}
\label{subsubsect:backg-max-prob}
\begin{figure}[t]
    \centering
    \begin{subfigure}[b]{0.3\textwidth}
      \centering
      \includegraphics[width=\textwidth]{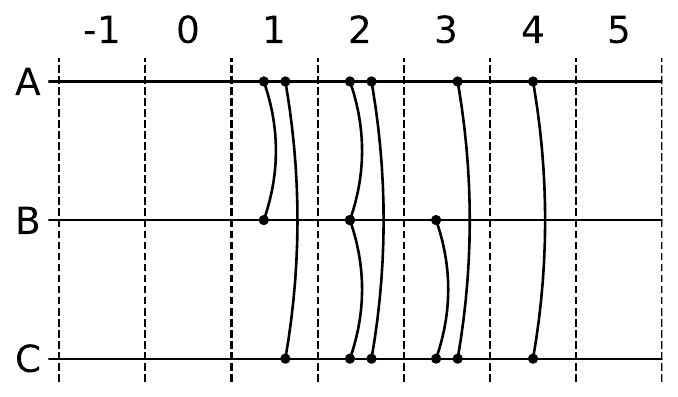}
      \caption{Link stream representation of temporal network}
    \end{subfigure}
    ~
    \begin{subfigure}[b]{0.3\textwidth}
      \centering
      \includegraphics[width=\textwidth]{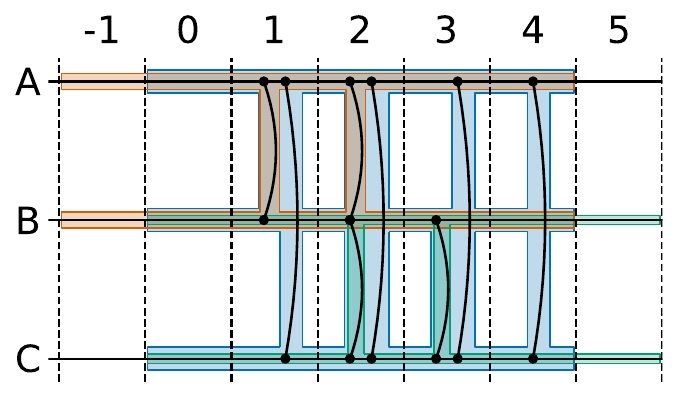}
      \caption{($\delta, \gamma$)-maximal cliques (for existing definition)}
    \end{subfigure}
    ~
    \begin{subfigure}[b]{0.3\textwidth}
      \centering
      \includegraphics[width=\textwidth]{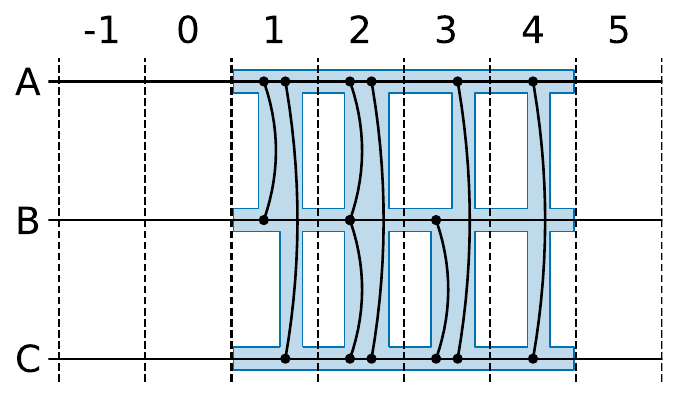}
      \caption{($\delta, \gamma$)-maximal cliques (for our definition)}
    \end{subfigure}
    \caption{Given $\delta = 4$ and $\gamma = 2$ and temporal network (a) shown as a link stream model, then (b) highlights the three maximal ($\delta, \gamma$)-cliques $C_1 = (\{A,B\}, [-1,4])$ in orange (with backslash pattern), $C_2 = (\{B,C\}, [0,5])$ in green (with cross pattern), and $C_3 = (\{A,B,C\}, [0,4])$ in blue (with forward slash pattern), that are by existing definitions ($\delta, \gamma$)-maximal. Note that the edges included in $C_1$ and $C_2$ are also all included in $C_3$. Therefore, $C_1$ and $C_2$ are effectively subgraphs of $C_3$.
    Finally, (c) shows the only remaining ($\delta, \gamma$)-maximal clique, when the time interval is bounded by the first and last temporal edge occurrences included. This clique ($C_4 = (\{A,B,C\}, [1,4])$) is found by our proposed definition (Def.~\ref{def:dgnew-clique}).
    Notice, that $C_4$ is the bounded form of $C_3$ from (b) and that the bounded forms of cliques $C_1$ and $C_2$ are no longer maximal.}
    \label{fig:failure}
\end{figure}
Although theoretically sound, the $\delta$-maximal clique and Banerjee \& Pal's ($\delta, \gamma$)-maximal clique definitions create a practical problem where some maximal cliques are in practice a strict subgraph of other maximal cliques. 
This problem originates from the fact that the definitions allow the time interval to be expanded beyond the scope of the temporal edge instances themselves. 
For example, consider a network consisting of a single temporal edge instance $(3, u, v)$. With $\delta$ set to 3, the $\delta$-clique $C = (\{u,v\}, [1, 5])$ is a valid and $\delta$-maximal clique, despite including a length 2 time interval before and after the only actual edge instance.
By allowing these empty intervals, the interpretation of a clique becomes more complex as one can not assume that the first and last interactions of the clique members actually occur at the borders of the time interval.
Moreover, the link stream model shown in Figure~\ref{fig:failure}b, demonstrates how empty intervals lead to the aforementioned practical problem of some maximal cliques being effectively subgraphs of others.

We combat this problem by adding bounds on the start and end of the time interval, determined by the actual temporal edge instances, to the ($\delta, \gamma$)-clique definitions of Banerjee \& Pal.
We additionally define $\gamma$ such that it describes a threshold on the cumulative weight of the temporal edge instances, instead of a threshold on the number of temporal instances.
As such, given our definition of frequency in section~\ref{subsect:backg-notation}, the following definition now holds both for weighted and unweighted networks.\vspace{5pt}
\begin{definition}[($\delta, \gamma$)-clique] \label{def:dgnew-clique}
    Given a time duration $\delta \in \mathbb{Z}+$ and a cumulative weight threshold $\gamma \in \mathbb{R}$, a ($\delta, \gamma$)-clique of the temporal network $\mathcal{G}$ is a vertex set and time interval pair $C = (X, [t_b, t_e])$ with $X \subseteq V_{\mathcal{G}}$ and $t_b,t_e \in T$ such that for all $u,v \in X$, where $v \neq w$, and $\tau \in [t_b, t_e-\delta+1]$, it holds that $f((u,v), [\tau, \tau + \delta - 1]) \geq \gamma$ and there exist temporal edge instances $(t_b, u, v)$ and $(t_e, u', v')$ with $u,v,u',v' \in X$.
\end{definition} \vspace{5pt}
Note that this also redefines $\delta$-cliques for a temporal network given $\gamma = 1$ and that the extension to $\delta$-maximal cliques follows definition~\ref{def:dmax-clique}.
The extension for ($\delta, \gamma$)-cliques to maximal cliques is subsequently similar to that of $\delta$-cliques. \vspace{5pt}

\begin{definition}[($\delta, \gamma$)-maximal clique]\label{def:dgmax-clique}
    A ($\delta, \gamma$)-clique $C = (X, [t_b, t_e])$ is considered \emph{($\delta, \gamma$)-maximal}, when there exists no ($\delta, \gamma$)-clique $C' = (X', [t'_b, t'_e])$ such that either $X \subset X'$, $t'_b \leq t_b$, and $t'_e \geq t_e$ holds or $X = X'$ and $t'_b < t_b$ or $t'_e > t_e$ holds.
\end{definition} \vspace{5pt}

\section{Fast \texorpdfstring{$(\delta,\gamma)$}{(delta,gamma)}-maximal clique enumeration}
\label{sect:methods}

In this section we propose a new $(\delta,\gamma)$-maximal clique enumeration algorithm.  
The overall methodology, described in Section~\ref{subsect:methods-overall}, is inspired by the state-of-the-art two-phase approach introduced by Banerjee \& Pal~\cite{banerjee2021two,banerjee2024two}.
In Section~\ref{subsect:methods-phase-one}, we introduce our approach to the initial so-called \emph{stretching phase}, which determines all 2-node duration-wise maximal $(\delta,\gamma)$-cliques.
Our contribution with respect to previous work lies in the fact that it accommodates weighted networks while also reducing the time complexity of this step from $\mathcal{O}(\gamma m)$ to $\mathcal{O}(m)$.
Then in Section~\ref{subsect:methods-phase-two}, we propose a new approach for the so-called \emph{bulking phase} which grows the node sets of the duration-wise maximal cliques.
This approach utilises already computed values to efficiently compute node set extensions and prune branches of the search tree, resulting in substantial speedups, as we will demonstrate in Section~\ref{sect:experiments}.

\subsection{Overall approach}
\label{subsect:methods-overall}
The state-of-the-art $(\delta,\gamma)$-maximal clique enumeration algorithm introduced by Banerjee \& Pal~\cite{banerjee2021two,banerjee2024two}, consists of two phases: (1) the stretching phase and (2) the shrink and bulk phase.
The stretching phase, takes, for each edge, the set of temporal instances as input and produces the set $C^s$ consisting of all 2-node duration-wise maximal $(\delta,\gamma)$-cliques.
In other words, it finds the maximally stretched time intervals of each edge that satisfy the $\delta$- and $\gamma$-constraints of a $(\delta,\gamma)$-clique.
Subsequently, the shrink and bulk phase continually expands the node sets (of the cliques in $C^s$) one node at a time, by combining the cliques in $C^s$.
As the process expands the node sets, it also shrinks the time interval when required to continue to conform to the $\delta$- and $\gamma$-constraints.
Through this node expansion process, the set of all $(\delta,\gamma)$-maximal cliques ($R$) is produced.

Our new proposed enumeration algorithm follows the same overall approach of first stretching the time intervals of individual edges before expanding the node sets.
However, our corrected definition of a $(\delta,\gamma)$-clique and the expansion to weighted networks, require us to propose entirely new (and faster) approaches for both phases.
For example, note that during the bulk phase, $(\delta,\gamma)$-cliques no longer strictly shrink their time interval during node expansions. 
After all, the time interval of a clique is now constrained by the actual temporal instances, whereas for Banerjee \& Pal~\cite{banerjee2021two,banerjee2024two} the time interval extended as far as the $\delta$- and $\gamma$-constraints remained satisfied, i.e., it extended to the \emph{maximum temporal growth} of the clique.
Thus, if temporal instances introduced by node expansions are within the constraints of the maximum temporal growth, the time interval can grow due to node expansion during our new bulk phase.
As such, for our proposed method, we refer to the second phase as the \emph{bulking phase}.

We describe our proposed approaches for the stretching and bulking phases in detail in Sections~\ref{subsect:methods-phase-one} and~\ref{subsect:methods-phase-two}, respectively.
The pseudo code describing our proposed methods are presented, respectively, in Algorithms~\ref{alg:stretch-wrapper},~\ref{alg:stretch-edge} and Algorithms~\ref{alg:bulk-wrapper}--\ref{alg:bulk-inner-recursion-overlap} in Appendix~\ref{app:A}. 
Lines throughout all algorithms are numbered sequentially. 
Hereafter, we refrain from referring to specific algorithms and only specify the relevant lines.
\subsection{Stretching phase}
\label{subsect:methods-phase-one}
The stretching phase initializes the $(\delta,\gamma)$-maximal clique enumeration process by first exploring the potential temporal growth for every edge. 
In other words, the stretching phase enumerates all 2-node duration-wise maximal $(\delta,\gamma)$-cliques.
It does so by evaluating the temporal instances of each edge separately (line~\ref{line:stretch-start}), while skipping any edges for which the cumulative weight of the instances is lower than $\gamma$ (line~\ref{line:stretch-start-weightcheck}).
The algorithm proposed by Banerjee \& Pal~\cite{banerjee2021two,banerjee2024two} for this phase is highly reliant on $\gamma$ indicating the number of occurrences of temporal edge instances. 
Since we can not rely on this for weighted networks, we instead propose a new (repeated) three step process (see lines~\ref{line:stretch-edge-start}--\ref{line:stretch-edge-end}): (1) initialization; (2) discovery of the start of a valid ($\delta,\gamma$)-clique; and (3) finding the end of said clique such that it is duration-wise maximal. 
During this process, we continually increase two indices ($bi$ and $ei$) on the temporal instances of the edge, such that the indicated instances are never more than $\delta$ timestamps apart. 
By keeping track of the current cumulative weight of the range of instances specified by those indices (\emph{current\_weight}), we can find the set of all 2-node duration-wise maximal $(\delta,\gamma)$-cliques of the given edge. 

\begin{figure}[t]
    \setlength\tabcolsep{4pt}
    \begin{minipage}[t]{0.49\linewidth}
        \centering
        \includegraphics[width=.8\textwidth]{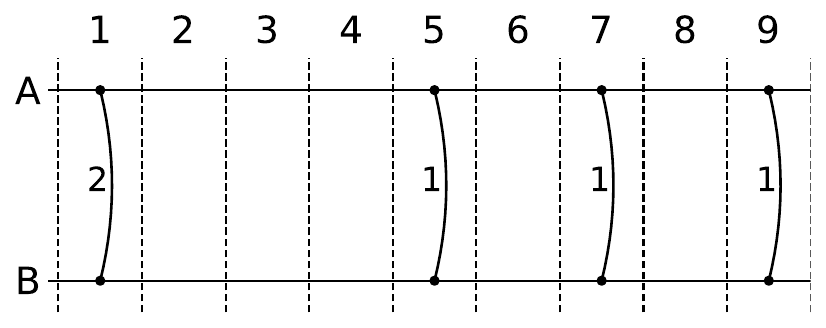}
        \begin{tabular}[b]{|ccccc|c|l|} \hline
            $bi$ & $t_{bi}$ & $ei$ & $t_{ei}$ & $cw$ & $C^s$ & step(s) \\ \hline
            0 & 1 & 0 & 1 & 2 & $\{\}$ & 1 \\
            0 & 1 & 1 & 5 & 3 & $\{\}$ & 2 \\
            0 & 1 & 1 & 5 & 3 & $\{\}$ & 3a, 3b \\
            1 & 5 & 1 & 5 & 1 & $\{[1,5]\}$ & 3d(ii) \\
            1 & 5 & 2 & 7 & 2 & $\{[1,5]\}$ & 2 \\
            1 & 5 & 3 & 9 & 3 & $\{[1,5]\}$ & 2 \\
            1 & 5 & 3 & 9 & 3 & $\{[1,5]\}$ & 3a, 3b \\
            x & x & x & x & x & $\{[1,5],$ & 3c \\ 
             &  &  &  &  & \ \ $[5,9]\}$ &  \\ \hline
        \end{tabular}
        \subcaption{$\delta = 5, \gamma = 3$}
    \end{minipage}
    \begin{minipage}[t]{0.49\linewidth}
        \centering
        \includegraphics[width=.8\textwidth]{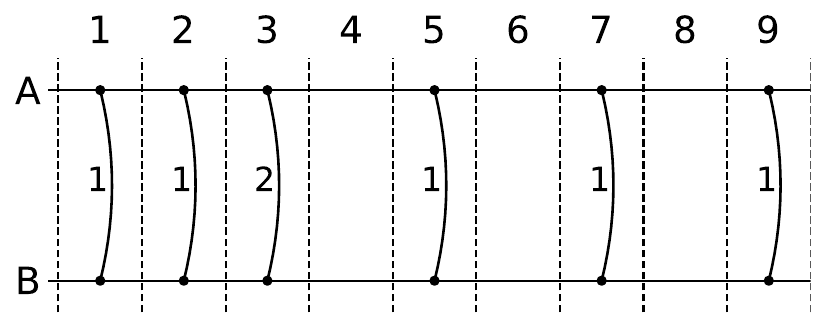}
        \begin{tabular}[b]{|ccccc|c|l|} \hline
            $bi$ & $t_{bi}$ & $ei$ & $t_{ei}$ & $cw$ & $C^s$ & step(s) \\ \hline
            0 & 1 & 0 & 1 & 1 & $\{\}$ & 1 \\
            0 & 1 & 1 & 2 & 2 & $\{\}$ & 2 \\
            0 & 1 & 2 & 3 & 4 & $\{\}$ & 2 \\
            0 & 1 & 3 & 5 & 5 & $\{\}$ & 3a \\
            1 & 2 & 3 & 5 & 4 & $\{\}$ & 3b \\
            1 & 2 & 3 & 5 & 4 & $\{\}$ & 3a, 3b \\
            2 & 3 & 4 & 7 & 4 & $\{\}$ & 3d(iii) \\
            2 & 3 & 4 & 7 & 4 & $\{\}$ & 3a, 3b \\
            3 & 5 & 4 & 7 & 2 & $\{[1,7]\}$ & 3d(ii) \\
            3 & 5 & 5 & 9 & 3 & $\{[1,7]\}$ & 2 \\
            x & x & x & x & x & $\{[1,7]\}$ & 2 \\ \hline
        \end{tabular}
        \subcaption{$\delta = 5, \gamma = 4$}
    \end{minipage}
    \caption{The stretching process for two example link stream models. Edge weights are depicted alongside each edge, $cw$ indicates the \emph{current\_weight}, and ``step'' indicates which step in the process produced the values presented in that row. Note that, for the cliques added to $C^s$, we only show their time interval in the tables.}
    \label{fig:stretch-examples}
\end{figure}

Below we describe each step of the stretching process and how they ensure that exclusively the set of all duration-wise maximal $(\delta,\gamma)$-cliques is added to result set $C^s$.
We demonstrate the execution of the process through two examples in Figure~\ref{fig:stretch-examples}.
\begin{enumerate}
    \item First, the indices $bi$ and $ei$, i.e., the begin and end index respectively, are initialized to point to the first temporal edge instance in the temporal ordering. Furthermore, the \emph{current\_weight} is initialized to the weight of this first instance (lines~\ref{line:stretch-11},~\ref{line:stretch-12}). 
    
    Note that during the remainder of the process, whenever $bi$ and $ei$ are increased, the \emph{current\_weight} is respectively decreased and increased accordingly.
    \item Second (lines~\ref{line:stretch-14}--\ref{line:stretch-24}), $ei$ is increased until the \emph{current\_weight} is greater or equal to $\gamma$.
    If at any point in this process the $\delta$-constraint (i.e., $t_{ei} - t_{bi} < \delta$) is no longer satisfied, $bi$ is increased until the constraint is satisfied once more (lines~\ref{line:stretch-18}--\ref{line:stretch-21}). 
    Additionally, if $ei$ ever exceeds the number of temporal edge instances, we know that no more valid ($\delta,\gamma$)-cliques can be found and the process is stopped (line~\ref{line:stretch-16}). 
    
    Once the \emph{current\_weight} is greater or equal to $\gamma$, we know that there exists a valid $(\delta,\gamma)$-clique (according to Definition~\ref{def:dgnew-clique}) with $t_{bi} \Rightarrow t_b$.
    Additionally, $t_{ei}$ now indicates the first timestamp from $t_{bi}$ that $\gamma$ weight is reached.
    Because knowledge of this timestamp plays a vital role for determining temporal overlap between cliques in the bulking phase, it is stored as a clique property ($t_{ei} \Rightarrow \mathit{tbMax}$).
    \item Finally, in order to find the duration-wise maximal clique whose time interval starts at $t_b$, the indices are systematically increased to ensure that the $(\delta,\gamma)$-clique with time interval [$t_b, t_{ei}$] always remains valid (lines~\ref{line:stretch-25}--\ref{line:stretch-60}).
    This is accomplished by ensuring that the cumulative weight of the instances in time interval [$t_{bi}, t_{ei}$] always remains at least $\gamma$ and that $\delta$-constraint (i.e., $t_{ei} - t_{bi} < \delta$) always holds.        
    To this end, the following steps are (repeatedly) performed:  

    \begin{enumerate}
        \item First, $ei$ is increased as long as doing so does not compromise the $\delta$-constraint (lines~\ref{line:stretch-28}--\ref{line:stretch-31}).
        Since at this point a minimum $\gamma$ weight is guaranteed for each $\delta$-time interval up to [$t_{bi}, t_{bi} + \delta -1$], $ei$ can be safely increased as long as $t_{ei}$ does not exceed $t_{bi} + \delta -1$, i.e., as long as the $\delta$-constraint is satisfied.
        \item Next, $bi$ is continually increased as long as doing so does not drop the \emph{current\_weight} below $\gamma$, i.e., as long as the $\gamma$-constraint is satisfied (lines~\ref{line:stretch-32}--\ref{line:stretch-35}).
        \item If $ei$ now points to the last temporal edge instance, no further extensions to the clique time interval are possible.
        Therefore, the $(\delta,\gamma)$-clique with $t_{ei} \Rightarrow t_e$ is duration-wise maximal and is added to result set $C^s$.
        Since no further temporal edge instances exist, the process is subsequently stopped (lines~\ref{line:stretch-36}--\ref{line:stretch-39}).

        Step 3b ensured that $t_{bi}$ now indicates the last timestamp from which $\gamma$ weight can be reached in the clique interval.
        Similar to $\mathit{tbMax}$ in step 2, knowledge of this timestamp plays a vital role during the bulking phase and therefore $t_{bi} \Rightarrow \mathit{teMin}$ is stored as a clique property.  
        \item If the process was not stopped in step 3c, we next consider two possible situations. 
        Either the indices $bi$ and $ei$ were updated during steps 3a and 3b, or they were not.
        If they were updated, we continue our process by returning to step 3a. 
        On the other hand, if the indices remained unchanged through steps 3a and 3b, we must check whether any further growth is possible (lines~\ref{line:stretch-41}--\ref{line:stretch-58}). 
        In other words, we must check if growth to the earliest time interval for which the $\gamma$ minimum weight is not yet guaranteed, i.e., [$t_{bi}+1, t_{bi} + \delta$], is possible.
        If not, we know that the ($\delta,\gamma$)-clique with time interval [$t_b, t_{ei}$] is duration-wise maximal.
        
        Because $bi$ was not updated in step 3b, we know that the cumulative weight for time interval [$t_{bi}+1, t_{bi} + \delta - 1$] is below $\gamma$.
        Thus, it need only be checked that a temporal edge instance exists at timestamp $t_{bi} + \delta$ and that it has sufficient weight.
        This leaves us with three possible scenarios, listed below. 
        \begin{enumerate}
            \item First, there may exist edge instance(s) at $t_{bi} + \delta$, but with insufficient weight. 
            In this scenario a new duration-wise maximal $(\delta,\gamma)$-clique is added to $C^s$ based on the same indices described in step 3c.
            If there are no further temporal edge instances beyond $t_{bi} + \delta$ then the process is stopped (lines~\ref{line:stretch-45}--\ref{line:stretch-48}).
            Otherwise (lines~\ref{line:stretch-49}--\ref{line:stretch-54}), further duration-wise maximal $(\delta,\gamma)$-cliques may exist. 
            To find these, the search process is resumed at step 2 with $bi = bi + 1$, $ei$ set to the last temporal edge instance considered during the check (i.e., the last instance up to and including timestamp $t_{bi} + \delta$), and the \emph{current\_weight} set according to the range of instances denoted by these indices. 

            Note that we may presume that the earliest possible new duration-wise maximal $(\delta,\gamma)$-clique starts at edge instance $bi + 1$. 
            After all, if this were not the case, then such a clique would need to include instance $bi$ and the time interval [$t_{bi}, t_{bi} + \delta - 1$] would need to have sufficient weight. 
            However, we have previously established this not to be the case.
            Additionally, since we also know that at this point $t_{bi+1} \geq t_{bi} + 1$ and $t_{ei} \leq t_{bi} + \delta$, it follows that $t_{ei} - t_{bi+1} < \delta$. 
            Therefore, it is guaranteed that the time interval [$t_{bi+1}, t_{ei}$] would be included in any duration-wise maximal $(\delta,\gamma)$-clique starting at $t_{bi+1}$.             
            Thus, by continuing our search from these indices, we guarantee future cliques are maximal on the left side of the time interval and that we do not miss any duration-wise maximal $(\delta,\gamma)$-clique.
            In other words, it guarantees we find exclusively the set of all 2-node duration-wise maximal $(\delta,\gamma)$-cliques.
            
            \item Second, the next temporal edge instance may be after $t_{bi} + \delta$ (lines~\ref{line:stretch-49}--\ref{line:stretch-54}). 
            Again, in this scenario a new duration-wise maximal $(\delta,\gamma)$-clique is added to $C^s$ based on the same indices described in step 3c and the search process is resumed at step 2 with the indices set as described above for scenario (i).
            \item Third, there may exist edge instance(s) at $t_{bi} + \delta$ with sufficient weight. 
            In other words, the clique time interval can be extended to include $t_{bi} + \delta$. 
            To check for further extensions, the search process is resumed at step 3a with variables updated according to the observed valid extension (which also align with those described above for scenario (i)).
        \end{enumerate} 
    \end{enumerate}
\end{enumerate}

Each loop of our stretching algorithm increases at least one of the two indices in each iteration.
Therefore, in the worst case we require a combined $2m$ iterations. 
Furthermore, given that all operations are themselves $\mathcal{O}(1)$, the time complexity of our proposed algorithm is $\mathcal{O}(2m) \to \mathcal{O}(m)$.
This improves upon Banerjee \& Pal's~\cite{banerjee2021two,banerjee2024two} stretching phase complexity of $O(\gamma m)$
Thus our proposed approach is more efficient and is able to process both weighted and unweighted networks.

\begin{figure}[t]
    \centering
    \includegraphics[width=0.8\textwidth]{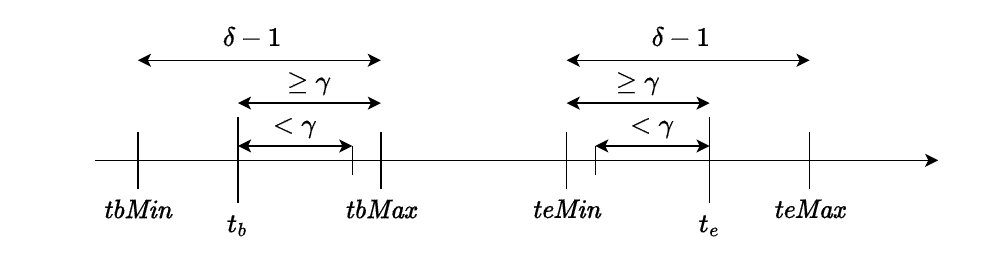}
    \caption{A visual representation of the clique time interval and properties on a timeline}
    \label{fig:clique-borders}
\end{figure}

Note that $\mathit{tbMax} -\delta + 1$ and $\mathit{teMin} +\delta - 1$ are equal to respectively the $t_b$ and $t_e$ properties used in the existing  temporal clique definitions (see Definition~\ref{def:d-clique}). 
We will refer to these clique properties in the remainder of this work as \emph{outer borders} and $\mathit{tbMin}$ and $\mathit{teMax}$, respectively.
See Figure~\ref{fig:clique-borders} for a visual representation of the various timestamp properties now defined for each duration-wise maximal $(\delta,\gamma)$-clique in $C^s$. 
\subsection{Bulking phase}
\label{subsect:methods-phase-two}
Following the stretching phase, the set $C^s$ now contains all 2-node duration-wise maximal $(\delta,\gamma)$-cliques.
During the bulking phase, we recursively expand the node sets of these cliques while updating their time intervals to remain duration-wise maximal, such as to find the set of all $(\delta,\gamma)$-maximal cliques ($R$).
Our proposed bulking method is entirely independent of $\gamma$, meaning that the value of $\gamma$ does not affect the time complexity of either phase.
Furthermore, the bulking phase's independence of $\gamma$, means that no special accommodations need to be considered for weighted networks.
We improve upon the algorithm introduced by Banerjee \& Pal~\cite{banerjee2021two,banerjee2024two}, by introducing (efficient) pruning of branches that will not include any new $(\delta,\gamma)$-maximal cliques. 
Additionally, by following a predefined search order, determined by the node labelling, we are able to further improve our pruning efficiency.
Moreover, it allows us to easily prune duplicate search branches, thereby improving on the space complexity by no longer requiring intermediate storage of all cliques of given sizes.
Our proposed method has been designed such that the pruning can be computed based almost entirely on values computed for the basic recursive expansion.

We first describe the recursive clique expansion loop, duplicate branch pruning and the effective re-use of computed cliques in Section~\ref{subsubsect:methods-recursion}.
Next, in Section~\ref{subsubsect:methods-overlap} we explain how we determine the duration-wise maximal time interval overlap resulting from combining three known cliques for node expansions. 
Then, we discuss how we efficiently determine which branches may be pruned in Section~\ref{subsubsect:methods-cuts}.
Finally, in Section~\ref{subsubsect:methods-searchorder}, we argue how our chosen node labeling scheme may improve the efficiency of our approach, for example, by increasing the number of pruned branches.
\subsubsection{Recursive search and preventing duplicate branches}
\label{subsubsect:methods-recursion}
The bulking process consists, at its basis, of recursive single-node extensions to the node sets of the duration-wise maximal $(\delta,\gamma)$-cliques in $C^s$.
Such a simple recursive process would however end up in many duplicate states/branches.
In order to prevent us from processing such duplicate states, we only continue the recursive search for clique extensions where the new node has a higher label than all nodes in the cliques' current node set (see lines~\ref{line:rec-conds-outer} and~\ref{line:rec-conds-inner}). 
As such, node set $\{a,b,c,d\}$ can only be found by extending the clique with node set $\{a,b\}$ to $\{a,b,c\}$ and then $\{a,b,c,d\}$.
Therefore, each state is visited only one time and no duplicate work is performed.

Due to slight differences for the node extension (and pruning) logic, the recursive process is split between an outer-recursion step (Algorithms~\ref{alg:bulk-outer-recursion},~\ref{alg:bulk-outer-recursion-overlap}) and an inner-recursion loop (Algorithms~\ref{alg:bulk-inner-recursion},~\ref{alg:bulk-inner-recursion-overlap}).
In the outer-recursion step, we determine all extensions from a 2-node clique ($C$) to 3-node cliques based on their shared neighbors (lines~\ref{line:outer-ext-start}--\ref{line:outer-ext-end}).
Since this essentially extends a single edge ($u,v$) to a triangle ($u,v,w$) of three edges, we can perform this extension entirely based on the duration-wise maximal cliques already in $C^s$.
Given sufficient time interval overlap, we combine cliques $C, C_u, C_v \in C^s$, with node sets $\{u,v\}$, $\{u,w\}$, and $\{v,w\}$, while ensuring that the resulting 3-node cliques remain duration-wise maximal (see lines~\ref{line:outer-overlap-u-select}--\ref{line:outer-overlap-interval-select-end}).

Subsequently, the inner-recursion loop extends at each step a clique $(C)$ by one new node based on the extensions $(C_p \in E_{prev})$ that were computed in the previous recursive step (lines~\ref{line:inner-ext-start}--\ref{line:inner-ext-end}).
We may rely on extensions from the previous recursive step since, for an extension from $\{a,b,c\}$ to $\{a,b,c,d\}$ to be possible, we must have also determined a valid extension to $\{a,b,d\}$ in the previous step (see Figure~\ref{fig:extension-example} for a visual representation).
Thus, the inner-recursion loop attempts to combine valid extensions from the previous step with the clique currently under consideration.
Note, that the only edge of the new extension node set ($C.X \cup C_p.X$) not yet covered by cliques $C$ and $C_p$, is the edge connecting their respective extensions, i.e, the edge connecting nodes $\max(C.X)$ and $\max(C_p.X)$ (the dotted line in Figure~\ref{fig:extension-example}).
Thus, given sufficient time interval overlap, the inner recursion loop combines cliques $C$, $C_p \in E_{prev}$, and $C_e \in C^s$, where $C_e.X = \{\max(C.X),\max(C_p.X)\}$, while again ensuring that the resulting cliques remain duration-wise maximal (see lines~\ref{line:inner-overlap-u-select}--\ref{line:inner-overlap-interval-select-end}).

Thus, through both the outer- and inner-recursive steps, we ensure that all considered cliques are duration-wise maximal.
To establish whether a clique ($C$) is also $(\delta,\gamma)$-maximal, we check that it is not temporally dominated by any of its extended cliques $C_{new}$ (see lines~\ref{line:outer-check-maximal-start}--\ref{line:outer-check-maximal-end} and~\ref{line:inner-check-maximal-start}--\ref{line:inner-check-maximal-end}). 
We say that $C$ is \emph{temporally dominated} by $C_{new}$ when $C_{new}.t_b \leq C.t_b$ and $C.t_e \leq C_{new}.t_e$ (see also lines~\ref{line:tempdom-start}--\ref{line:tempdom-end}). 
If no such node extension exists, we know that $C$ is a $(\delta,\gamma)$-maximal clique according to Definition~\ref{def:dgmax-clique}.
In such cases, we store the clique $C$ in result set $R$, such that at the end of our search process, $R$ contains all $(\delta,\gamma)$-maximal cliques (see lines~\ref{line:outer-save-maximal-start}--\ref{line:outer-save-maximal-end} and~\ref{line:inner-save-maximal-start}--\ref{line:inner-save-maximal-end}).

Note that both the outer- and inner-recursion steps require combining three known cliques. 
Thus, the recursion effectively re-uses already computed cliques.
To combine three cliques we need to determine their duration-wise maximal time interval overlap.
We describe the logic and time complexity of this process in Section~\ref{subsubsect:methods-overlap} below.
\subsubsection{Duration-wise maximal time interval overlap of three cliques}
\label{subsubsect:methods-overlap}
\begin{figure}[t]
    \begin{minipage}[b]{0.2\textwidth}
        \centering
        \includegraphics[width=\textwidth]{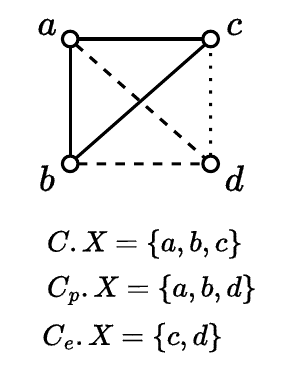}
        \caption{Inner-recursion extension visualization}
        \label{fig:extension-example}
    \end{minipage}
    \begin{minipage}[b]{0.78\linewidth}
        \centering
        \includegraphics[width=\textwidth]{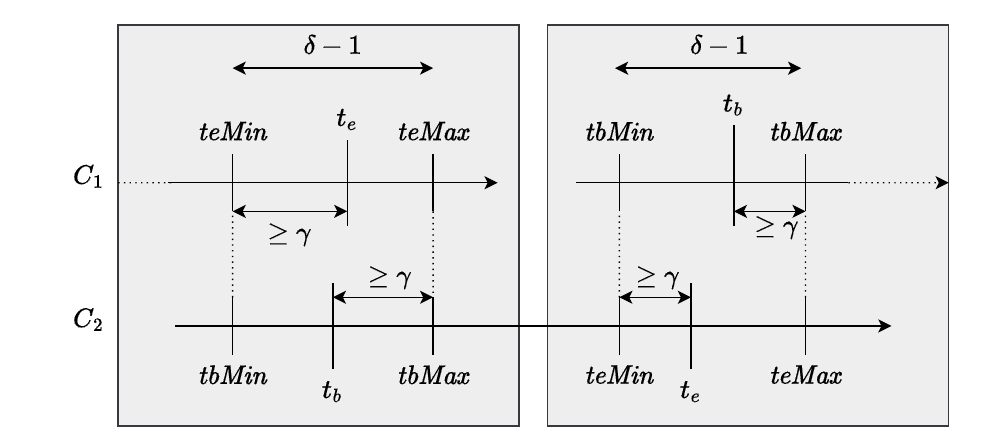}
        \caption{Time interval overlap extremes}   
        \label{fig:clique-overlap-extremes}
    \end{minipage}
\end{figure}
In the previous subsection, we described how the recursive process relies at each step on determining the duration-wise maximal time interval overlap between three cliques.
It is therefore crucial to establish a method of efficiently computing such an overlap.
Let us refer to the three cliques under consideration as $C_1$, $C_2$, and $C_3$.
We propose a five step approach to computing the duration-wise maximal time interval overlap. 

First, we determine if there exists a valid time interval overlap for $C_1$ and $C_2$. 
For any overlap between two time intervals to exist, we know that the left and right borders of one must be before and after the respective right and left borders of another. 
However, we want to check not only whether there is an overlap but also ensure that the overlap is sufficient such that at least $\gamma$ weight is included for all edges in the respective node sets.
Here, the previously introduced properties $\mathit{tbMin}$, $\mathit{tbMax}$, $\mathit{teMin}$, and $\mathit{teMax}$ provide our solution (see Figure~\ref{fig:clique-borders} for a reminder of how these properties relate to one another). 
We determine if a valid overlap exists by checking whether \[ C_1.\mathit{teMax} \geq C_2.\mathit{tbMax} \text{ and } C_1.\mathit{tbMin} \leq C_2.\mathit{teMin}.\]

\begin{proof}
Figure~\ref{fig:clique-overlap-extremes} shows the extreme cases where $C_1.\mathit{teMax} = C_2.\mathit{tbMax}$ or $C_1.\mathit{tbMin} = C_2.\mathit{teMin}$. 
Note that both $\mathit{tbMin}$ and $\mathit{teMax}$ indicate the extent to which the time interval may be freely grown.
Therefore, our overlap time intervals in these extreme cases would become $[C_1.\mathit{teMin}, C_2.\mathit{tbMax}]$ or $[C_1.\mathit{tbMin}, C_2.\mathit{teMax}]$. 
As shown in Figure~\ref{fig:clique-overlap-extremes}, the edges for both $C_1.X$ and $C_2.X$ are guaranteed at least $\gamma$ minimum weight for these time intervals.
For non-extreme cases, it then follows that the overlap time intervals will be at least of equal ($\delta$) length.
After all, even if one of the time intervals themselves is less than $\delta$ in length, their potential growth is up to at least $\delta$ length due to how $\mathit{tbMin}$ and $\mathit{teMax}$ are defined.
Given that we know that, for both $C_1$ and $C_2$, every $\delta$ duration has a minimum $\gamma$ weight for each edge in their respective node sets, this is guaranteed to also hold for the overlap time interval for all non-extreme cases.
\end{proof}

Second, having determined that there exists a valid overlap between cliques $C_1$ and $C_2$, we must next determine their shared outer borders ($\mathit{tbMin^{new}},\mathit{teMax^{new}}$). 
In other words, we determine the bounds on their combined potential time interval growth. 
Since the time interval overlap is defined for the combined set of edges from $C_1$ and $C_2$, the potential growth of the time interval overlap is subject to the limitations on this growth for both $C_1$ and $C_2$.
Thus, we take the maximum $\mathit{tbMin}$ and the minimum $\mathit{teMax}$, i.e., we shrink the borders to comply with both limitations.

In the third and fourth steps, we essentially repeat the first two steps but now for respectively the time interval of the overlap of $C_1, C_2$ and the time interval of $C_3$.
Following our earlier proof, this checks whether there exists an overlap between cliques $C_1$, $C_2$, and $C_3$ such that all edges defined by their respective node sets have at least $\gamma$ weight.
Furthermore, the time interval $[\mathit{tbMin^{new}},\mathit{teMax^{new}}]$ now determines the maximum time interval for which this holds.
These first four steps are described in our algorithms on respectively lines~\ref{line:outer-overlap-u-select}--\ref{line:outer-outborder-selected-end} and~\ref{line:inner-overlap-u-select}--\ref{line:inner-outborder-selected-end}.

\begin{figure}[t]
    \centering
    \includegraphics[width=0.9\textwidth]{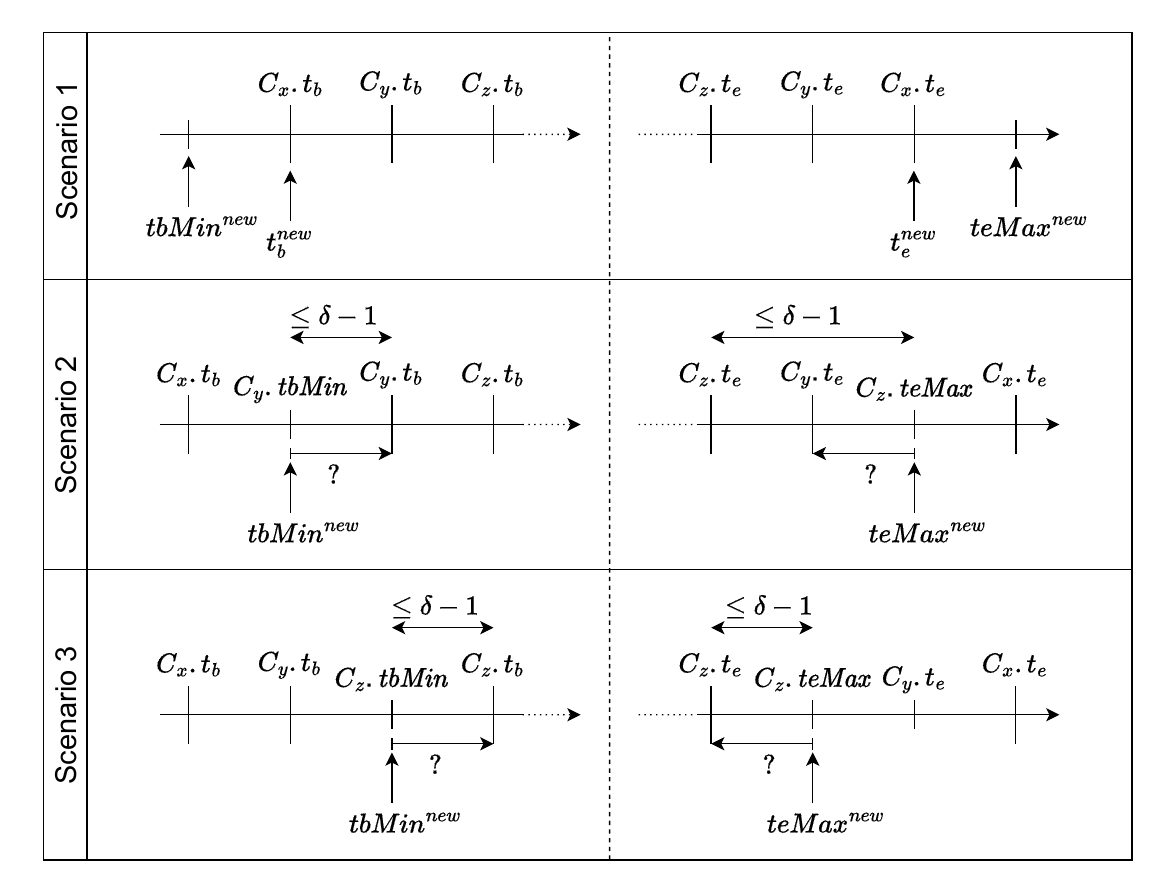}
    \caption{Time interval border selection scenarios}
    \label{fig:clique-border-select}
\end{figure}

For the final fifth step, we now need to determine the time interval overlap borders ($t_b$ and $t_e$) within time interval $[\mathit{tbMin^{new}},\mathit{teMax^{new}}]$, such that the resulting clique is duration-wise maximal.
Recall that we already know that any time interval within $[\mathit{tbMin^{new}},\mathit{teMax^{new}}]$ will constitute a valid $(\delta,\gamma)$-clique. 
Therefore, the first and last edge occurrences, for any of the edges in the merged node set, within this time interval, constitute the duration-wise maximal time interval borders.
In order to find these, we rely on the knowledge that each of the cliques under consideration ($C_1$, $C_2$, and $C_3$) are themselves duration-wise maximal, i.e., we rely on the fact that none of their edges have an occurrence before and after their respective $t_b$ and $t_e$ within this interval.
This leads us to consider three possible scenarios, as depicted in Figure~\ref{fig:clique-border-select}.

The first scenario is also the simplest. 
Here, all three $t_b$ occur after $\mathit{tbMin^{new}}$ or all three $t_e$ occur before $\mathit{teMax^{new}}$, i.e., they are all within the overlap time interval. 
Hence, all if-statements on lines~\ref{line:outer-overlap-interval-select-start}--\ref{line:outer-overlap-interval-select-end} and~\ref{line:inner-overlap-interval-select-start}--\ref{line:inner-overlap-interval-select-end} evaluate to ``False''.
As such, we can simply choose the respective minimum and maximum values of $t_b$ and $t_e$. 

The second and third scenarios are more complex.
Here, one or more $t_b$ or $t_e$ occur outside the time interval $[\mathit{tbMin^{new}},\mathit{teMax^{new}}]$.
In these cases, we can not fully rely on previously computed values.
Instead, we must determine the earliest (or latest) edge occurrence in the time intervals depicted with a `$?$' in Figure~\ref{fig:clique-border-select}. 
In these scenarios, one or two of the if statements on lines~\ref{line:outer-overlap-interval-select-start}--\ref{line:outer-overlap-interval-select-end} and~\ref{line:inner-overlap-interval-select-start}--\ref{line:inner-overlap-interval-select-end} must be processed.
In other words, for the cliques whose $t_b$ or $t_e$ are outside the time interval we must determine the earliest edge occurrence, e.g., $t_b^x = \min(\{t | (t,u,v) \in E^{\mathcal{T}} \wedge u,v \in C_x.X \wedge t \geq \mathit{tbMin^{new}}\}$.

Given cliques that represent a single edge, we can determine the earliest (or latest) timestamp by performing a binary search over the edge occurrences of said edge. 
Such a binary search has time complexity $\mathcal{O}(\log(k))$, where $k$ is the (average) number of temporal occurrences of an edge.
For the outer-recursion step (lines~\ref{line:outer-overlap-interval-select-start}--\ref{line:outer-overlap-interval-select-end}), which deals solely with cliques representing a single edge, this leads to a time complexity of $\mathcal{O}(\log(k))$ for scenario 2 and $\mathcal{O}(2\times\log(k))$ for scenario 3.

For the inner-recursion (lines~\ref{line:inner-overlap-interval-select-start}--~\ref{line:inner-overlap-interval-select-end}), we may be dealing with more edges.
Worst case we may need to check $\frac{|X_{new}|\times(|X_{new}|-1)}{2} - 1$ edges.
This leads to a worst case time complexity of $O(2 \times (\frac{|X_{new}|\times(|X_{new}|-1)}{2} - 1) \times \log(k))$. 
As such, determining the new $t_b$ and $t_e$ of the time interval overlap, is by far the most expensive operation, in terms of time complexity, during our recursive expansion of cliques. 
However, note that we may stop checking edges if at any point the earliest/latest occurrence found corresponds to the outer border.
Especially for temporal networks that are periodic, such as face-to-face networks where edges are created from periodic measurements (e.g., every 60 seconds), this allows us to often stop early, resulting in much better performance.

Following the fifth step that determined the new $t_b$ and $t_e$, we can now define and return our new duration-wise maximal overlap clique(s) on lines~\ref{line:outer-overlap-finish-start}--\ref{line:outer-overlap-finish-end} and~\ref{line:inner-overlap-finish-start}--\ref{line:inner-overlap-finish-end}. 

\subsubsection{Efficient branch pruning}
\label{subsubsect:methods-cuts}
In section~\ref{subsubsect:methods-recursion} we explained that we prevent our recursive search from processing duplicate branches by disallowing node extensions to nodes with a smaller label than the current maximum label in the node set.
As a direct consequence, many branches of the search tree no longer include any $(\delta,\gamma)$-maximal cliques.
After all, when a node extension exists to a smaller labelled node such that the current state is not maximal, this inevitably prevents us from reaching the maximal clique from that branch.
Here, we discuss how we may efficiently identify and prune such branches.

\begin{figure}[t]
    \centering
    \begin{subfigure}[t]{0.2\textwidth}
        \centering
        \includegraphics[width=\textwidth]{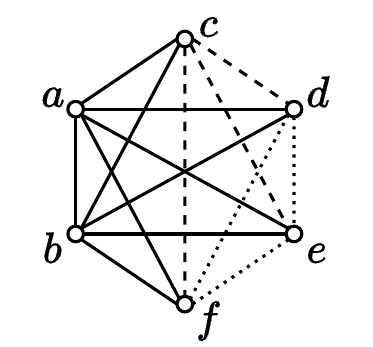}
        \caption{Static graph}
        \label{fig:cut-example-full}
    \end{subfigure} 
    ~
    \begin{subfigure}[t]{0.3\textwidth}
        \centering
        \includegraphics[width=.66\textwidth]{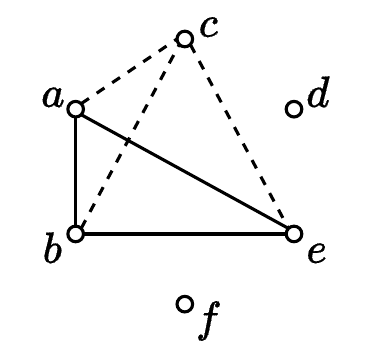}
        \caption{Check whether the root state is temporally dominated by inclusion of node $c$}
        \label{fig:cut-example-ext-root}
    \end{subfigure}
    ~
    \begin{subfigure}[t]{0.45\linewidth}
        \centering
        \includegraphics[width=.44\textwidth]{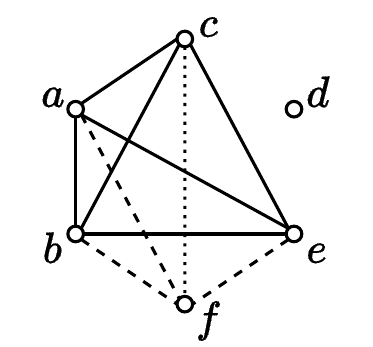}
        \caption{Check whether extension to $f$ is temporally dominated by considering shared (dashed) and unique (dotted) edges} 
        \label{fig:cut-example-ext}
    \end{subfigure}
    \caption{Bulking phase branch cut example}
\end{figure}

For a branch to be pruned, it must be proven that neither the root state of the branch nor any of its sub-states (i.e., all recursively reachable child-states) are $(\delta,\gamma)$-maximal. 
Specifically, given that all states in our search tree are duration-wise maximal, we only need to show that all sub-states are temporally dominated by another state with a larger node set.
The vast majority of information required to show that a branch can be pruned, is obtained as a natural part of the recursive process in \emph{neigbhoring branches}.
Neighboring branches are those whose root states have the same parent state.
For example, in Figure~\ref{fig:cut-example-full}, states $\{a,b,c\}$, $\{a,b,d\}$, $\{a,b,e\}$, and $\{a,b,f\}$ are neighboring branches with parent state $\{a,b\}$.
Note that for the outer recursion step, which processes single links, neighboring branches are considered to be all cliques connecting to the shared neighbors of the link.
Thus, in Figure~\ref{fig:cut-example-full}, when processing $\{a,b\}$ the neighboring branches are $\{a,c\}$, $\{b,c\}$, $\{a,d\}$, $\{b,d\}$, etc.
Furthermore, note that only neighboring branches with smaller maximum node labels, may provide the evidence needed to cut a branch. 
After all, only branches with smaller maximum node labels may include additional nodes.
For example, for state $\{a,b,e\}$, only $\{a,b,c\}$ and $\{a,b,d\}$ may include larger nodes sets due to the inclusion of nodes $c$ and $d$, which $\{a,b,e\}$ may no longer include.
As such, to determine whether branch $\{a,b,e\}$ can be pruned, we should process $\{a,b,c\}$ and $\{a,b,d\}$ first. 
Therefore, we enforce a search order which processes child-states in order of increasing maximum node label.

Let us assume that we are dealing with a network whose static representation is the graph shown in Figure~\ref{fig:cut-example-full}.
Additionally, let us assume we are currently processing a clique ($C$) with node set $\{a,b,c\}$. 
Then, in Figure~\ref{fig:cut-example-full} the edges included in cliques in the current and neighboring branch states are shown as solid lines. 
Furthermore, the edges included by direct extensions to $C$ are shown as dashed lines and the remaining edges as dotted lines.
Using this as our running example, we explain how we determine whether the neighboring branch ($C_{b}$) with node set $\{a,b,e\}$ can be pruned.

First, we confirm that the root of the neighboring branch is not $(\delta,\gamma)$-maximal. 
Since extensions to the current state already need to be computed as part of the recursive search process, this can be confirmed efficiently.
After all, if, from our current state ($C$), an extension to $C_{new}$ with node set $\{a,b,c,e\}$ exists that temporally dominates the root state $C_{b}$ (see Figure~\ref{fig:cut-example-ext-root}), then we know that $C_{b}$ is not $(\delta,\gamma)$-maximal.
If no such extension exists, we can not determine at this stage whether the root state is $(\delta,\gamma)$-maximal and we do not prune the branch (see lines~\ref{line:outer-cut-newdom} and~\ref{line:inner-cut-newdom}).

Next, we need to ensure that all sub-states, i.e., all extensions, of $C_{b}$ are not $(\delta,\gamma)$-maximal. 
In our example, this means we must show that, applying the same extensions to $C_{b}$ and $C_{new}$, maintains the latter's temporal dominance.
To this end, we must first confirm whether the same extensions are possible.
Note that all edges unique to $C_{new}$ and its extensions, compared to $C_{b}$ and its extensions, are those connected to node $c$.
Figure~\ref{fig:cut-example-full} indicates that these unique edges are each covered by at least one of the single node extensions to the current state (dashed lines). 
Therefore, it is simple to establish that $C$ and $C_{b}$ have the same possible extensions, by confirming that an extension from $C$ exists for each node to which $C_{b}$ may be extended. 
Moreover, if $C_{new}$ also has sufficient time interval overlap with the \emph{minimum extension growth} of $C$, then we know that extensions exist from $C_{new}$ to every node that $C_{b}$ may be extended to. 
The minimum extension growth of $C$ is defined as the smallest outer borders of single node extensions to $C$ (i.e., $\max(E^r.\mathit{tbMin})$ and $\min(E^r.\mathit{teMax})$ with $E^r$ the set of all single-node extensions to $C$).
Consequently, it is then ensured that, as long as $C_{b}$'s extensions are temporally dominated, any larger (multi-)node extensions must either exist for both the current state and the branch root state or for neither.
As such, in our example we need only confirm that the extension to $\{a,b,c,f\}$ exists and has sufficient time interval overlap with $C_{new}$.
If this holds, we say that $C_{b}$ is \emph{spatial growth dominated} by $C_{new}$ (see lines~\ref{line:spatial-growth-start}--\ref{line:spatial-growth-end},~\ref{line:outer-cut-spatdom}, and~\ref{line:inner-cut-spatdom}).

Upon confirming that $C_{b}$ is spatial growth dominated, we next need to show that its extensions are also temporally dominated, i.e., that $C_{b}$ is \emph{temporal growth dominated}. 
We may determine this by analysing the impact on the potential growth of the time interval, by edges that are added through node extensions. 
Recall from Section~\ref{subsubsect:methods-overlap}, that the addition of edges, i.e, combining cliques, may either maintain, grow, or shrink the time interval ($[t_b, t_e]$).
Importantly, the potential growth of the interval borders is only limited by the outer borders ($\mathit{tbMin}$ and $\mathit{teMax}$) and the outer borders themselves may only shrink with edge additions.
Moreover, the time interval may only shrink if the outer border shrinkage restricts it as such (e.g., in scenarios 2 and 3 in Figure~\ref{fig:clique-border-select}). 

To determine whether $C_{b}$ is temporal growth dominated, we need to establish the effects of all edges with which it may be extended.
For our example, Figure~\ref{fig:cut-example-ext} shows the only possible extension to $C_{b}$, i.e., the extension to node $f$. 
Here, the newly added edges are depicted with dashed and dotted lines.
Specifically, the dotted lines show the edges that are uniquely added through extensions to $C_{new}$.
These unique edges are the only reason that the outer borders of $C_{new}$ may be shrunk further than the outer borders of $C_{b}$.
Note that each of these edges are included in at least one of the single-node extensions ($E^r$).
Therefore, the potential shrinkage of the outer borders of $C_{new}$ through extensions, is determined by the minimum extension growth of $C$. 
To determine whether the branch root state is temporal growth dominated, we thus check if its actual potential growth may exceed the minimum extension growth of $C$ (see lines~\ref{line:temporal-growth-start}--\ref{line:temporal-growth-end}).

The most obvious reason for the actual temporal growth of $C_{b}$ to not exceed the minimum extension growth of $C$, is for its outer borders to be temporally dominated by the minimum extension growth (see lines~\ref{line:cut-tg-1-start}--\ref{line:cut-tg-1-end}). 
If this is not the case, we must check that none of the potential extensions to $C_{b}$ have time interval borders that exceed the minimum extension growth (see lines~\ref{line:cut-tg-2-start}--\ref{line:cut-tg-3-end}).
Specifically, this means we must check the time intervals of cliques representing edge(s) with which $C_{b}$ may be extended (i.e., cliques representing the dashed edges in Figure~\ref{fig:cut-example-ext}).
These edges can be split in two categories: those covered by neighboring branches (e.g., $(a,f)$ and $(b,f)$ in $\{a,b,f\}$); and those connecting the extension nodes (e.g., edge $(e,f)$).
For the former, we select the relevant cliques, that have sufficient time interval overlap to be used in a potential extension to $C_{b}$, from those previously computed on lines~\ref{line:outer-pot-select} and~\ref{line:inner-pot-select}, and subsequently check whether their time intervals exceed the minimum extension growth on lines~\ref{line:cut-tg-2-start}--\ref{line:cut-tg-2-end}.
Only if this is not the case, do we check the latter edges, i.e., those connecting the node extension cliques (see lines~\ref{line:cut-tg-3-start}--\ref{line:cut-tg-3-end}).
Selecting relevant cliques from $C^s$ for these edges and checking them against the minimum extension growth, is the only part of our pruning process that requires us to consider cliques not essential to the recursive search at that time.
By checking these last, we minimize how often we need to consider `new' cliques.
If none exceed the minimum extension growth of $C$, than $C_{b}$ is, practically speaking, temporal growth dominated and the branch may be pruned.

In summary, we determine whether a neighboring branch may be pruned by confirming that its root state is temporally dominated, spatial growth dominated, and temporal growth dominated.
We can determine this by relying almost exclusively on information that is readily at hand in the current state of the recursive search.

\subsubsection{Custom node labeling for efficient search and pruning}
\label{subsubsect:methods-searchorder}
Here, we discuss how a custom node labeling can be used to improve the efficiency of the bulking phase.
Recall that the bulking process always starts from a 2-node clique.
As such, both the (potential) depth and width of the recursive search trees are dependent on the number of shared neighbors of the two nodes.
Furthermore, recall that our recursive process skips any extensions to nodes with lower labels.
Therefore, we can reduce the average depth and width of the recursive search trees by providing the nodes with the most neighbors the highest labels, thereby visiting them last.
Although we still visit each state exactly one time, regardless of which node labeling is used, there are benefits from the reduced width of the search trees for the pruning process.
Each node that a clique may be extended to adds conditions that must be checked and satisfied for a branch to be pruned.
Therefore, fewer potential extensions means fewer conditions to satisfy.
Additionally, this custom node labeling lends itself better to future parallelization.
After all, by reducing the width and depth of the most extreme search trees, the duration of processing each individual search tree, for the cliques in $C^s$, becomes more balanced.
Parallelization can utilise this balance, by dividing parts of the $C^s$ set over multiple processes with less risk of one process being excessively slower than the others.

Thus, in this work we use a custom node labeling based on the number of neighbors of a node, with fewer neighbors leading to a lower label.
\section{Data}
\label{sect:data}
For our experiments we collected a mixture of smaller network datasets used by previous works to evaluate their algorithms, as well as a new set of larger networks. 
Aside from their size, we selected the datasets such that (weighted) temporal cliques would have meaning within their respective contexts, i.e., so that clique results can be meaningfully interpreted. 
All datasets were pre-processed to sort by timestamp and exclude self-edges.
Resulting size statistics and some basic information on the networks, including their source, are provided in Table~\ref{tab:data-basics}.
\begin{table}[b]
    \caption{Summary of dataset statistics. Network type abbreviations: \emph{f2f} = face-to-face, \emph{com} = communication, \emph{trr} = trust rating, \emph{hyp} = hyperlink, and \emph{cli} = co-listening. The ``Wei.'' column indicates whether the dataset includes edge weights and the ``Per.'' column indicates which datasets are periodic.}
    \label{tab:data-basics}
    \setlength\tabcolsep{4pt}
    \begin{tabular}{l|rrr|lccc}
        Dataset & $|V|$ & $|E|$ & $|E^\mathcal{{T}}|$ & Lifetime & Type & Wei.  & Per.\\ \hline
        Hypertext~\cite{Isella:2011qo,sociopatterns}            &       113 &     2,196 &     20,818 & 2.5 days   & \emph{f2f} & No  & Yes \\
        College message~\cite{panzarasa2009patterns,snapnets}   &     1,899 &    13,838 &     59,835 & 193 days   & \emph{com} & No  & No \\
        Bitcoin~\cite{kumar2016edge,kumar2018rev2,snapnets}     &     5,881 &    21,492 &     35,592 & 5.21 years & \emph{trr} & Yes & No \\
        Infectious I~\cite{Isella:2011qo,sociopatterns}         &    10,972 &    44,517 &    415,912 & 80 days    & \emph{f2f} & No  & Yes \\
        Infectious II~\cite{Isella:2011qo,sociopatterns}        &       410 &     2,765 &     17,298 & 8h         & \emph{f2f} & No  & Yes \\ \hline
        Facebook-wosn-wall~\cite{viswanath2009evolution,konect} &    45,813 &   183,412 &    855,542 & 4.28 years & \emph{com} & No  & No \\
        Reddit hyperlinks~\cite{kumar2018community,snapnets}    &    67,180 &   309,667 &    858,488 & 40 months  & \emph{hyp} & Yes & No \\
        Enron~\cite{klimt2004enron,konect}                      &    86,978 &   297,456 &  1,134,990 & 4.5 years  & \emph{com} & No  & No \\
        Last.fm bands~\cite{konect,konstas2009social}           &   174,077 &   894,388 & 19,146,398 & 4.35 years & \emph{cli} & No  & No \\
        Last.fm songs~\cite{konect,konstas2009social}           & 1,084,602 & 4,413,377 & 19,150,606 & 4.35 years & \emph{cli} & No  & No \\
    \end{tabular}    
\end{table}

The \texttt{Hypertext}, \texttt{Infectious I}, and \texttt{Infectious II} datasets are periodic networks of face-to-face interactions between conference attendees. 
Each edge indicates that two conference attendees were in close physical proximity to one another at a given time, with measurements having occurred at specific intervals. 
Cliques in these networks may represent groups of attendees having an ongoing conversation. 
Temporal cliques specifically, may allow us to more accurately capture these groups by excluding people who happen to pass by the group at the time of one measurement. 

We also included several non-periodic datasets. 
Two of them capture co-listening of songs (\texttt{Last.fm songs}) and bands (\texttt{Last.fm bands}) by users of the Last.fm social network. 
Temporal cliques in these networks may indicate a lasting shared interest in the same songs or bands by groups of users.
Additionally, we included three communication networks, the \texttt{College message}, \texttt{Facebook-wosn-wall}, and \texttt{Enron} datasets.
Respectively, they capture the private messages sent on an online social network at the University of California, Facebook wall posts between online social media users, and email communication between employees of the Enron Corporation.
Temporal cliques in these networks may indicate friend groups or (Enron) departments, between whom more frequent communication with all members is to be expected.

Finally, to experiment with edge weighting, two weighted datasets were included. 
The first dataset captures the trust between Bitcoin users (\texttt{Bitcoin}) through ratings of one another.
The second dataset consists of hyperlinks between subreddits (\texttt{Reddit hyperlinks}), where weight determines the sentiment (positive or negative) that accompanied the source subreddit.
By requiring a minimum weight on the (temporal) cliques we can find sufficiently strongly connected groups of users and subreddits, where this is determined by the actual strength of the connections and not simply their frequency. 

\section{Experiments}
\label{sect:experiments}
In this section we present the experimental evaluation of our algorithm, in which we compare our algorithm to existing (state-of-the-art) methods. In the following subsections we discuss, respectively, the experimental setup and results. 
\subsection{Experimental setup}
\label{subsect:exp-setup}
As discussed in Section~\ref{sect:relwork}, several algorithms have been proposed for temporal clique enumeration since the first introduction of $\delta$-cliques.
We evaluated our proposed algorithm against each of them for the $\gamma = 1$ setting, i.e., the purely temporal setting.
Specifically, this means we compare against the algorithms introduced by Viard et al.~\cite{viard2016computing}, Himmel et al.~\cite{himmel2017adapting}, and the two-phase approach algorithm of Banerjee \& Pal.~\cite{banerjee2024two} that inspired our method.
For higher $\gamma$ values, we can only compare to the algorithm proposed by Banerjee \& Pal.~\cite{banerjee2024two}.
Since these implementations rely on the existing (non-instance-bounded) definitions of temporal cliques (see Section~\ref{subsect:backg-max}), a post-processing step is required to obtain the set of cliques that satisfies Definition~\ref{def:dgmax-clique}.
The implemented post-processing step is simple but inefficient.
As such, time required for the post-processing and storing of result cliques, are excluded from the runtimes we report. 
The reported runtimes do include the time required to read in the dataset.
As an additional consequence of the difference in definition, the $\delta$ value reported alongside the results is the one used by our algorithm, while the existing algorithms were executed with a value one lower.

For Banerjee \& Pal.~\cite{banerjee2024two}, we applied a minor fix to their implementation, which for larger $\gamma$ did not properly account for cases where the same link occurs multiple times at the same timestamp.
Subsequently, for all experiments, Himmel et al.~\cite{himmel2017adapting} and Banerjee \& Pal.~\cite{banerjee2024two} obtained, after post-processing, the same sets of cliques as our proposed algorithm.
The implementation of Viard et al.~\cite{viard2016computing} however, did not implement the exact periodicity phase discussed in Section~\ref{subsect:backg-max}.
Instead, at a $\delta$ representing exact periodicity, it provides the results of pre-periodicity.
Since we have chosen $\delta$ values that often correspond to exact periodicity, this means that the clique set provided by Viard et al.~\cite{viard2016computing} did not match.
Despite this and for the sake of a complete comparison with existing methods, we still report on their performance.

All experiments were performed on a server with 16 Intel Xeon E5-2630v3 cores and 512GB RAM.
However, note that all implementations, including that of our own algorithm, run sequentially and therefore only used a single thread.
Alongside runtimes, we report maximum space usage based on the python ``resource'' package using the \emph{ru\_maxrss} property.
Furthermore, note that due to hardware technicalities, space usage below 76MB is measured as 76MB.
Since cases with such low memory usage are of little interest, we simply report all such cases as ``$\leq76$''.
A 24 hour time limit and a 200GB memory usage limit were applied, before post-processing. Consequently, some results are missing and an indication of which limit was exceeded is shown.
All runtime and memory statistics reported in this paper were averaged over five runs.

The (modified) implementations of all algorithms can be found at \url{https://github.com/hdboekhout/Fast-delta-gamma-clique}.

\subsection{Results}
\label{subsect:exp-results}
For our experiments we consider the basic temporal setting in Section~\ref{subsubsect:results-temp}.
We then compare our proposed algorithm to that of Banerjee \& Pal~\cite{banerjee2024two} for larger $\gamma$ values in the unweighted setting in Section~\ref{subsubsect:results-unw}.
Finally, in Section~\ref{subsubsect:results-weighted} we compare the weighted and unweighted settings for our algorithm.
\subsubsection{The basic temporal setting}
\label{subsubsect:results-temp}
To evaluate the performance of our proposed algorithm, we first compare its performance against existing $\delta$-maximal clique enumeration algorithms, i.e., the basic temporal setting. 
For our algorithm, this means we enumerate $(\delta,1)$-maximal cliques in unweighted (versions of the) networks.
Each algorithm was applied to each dataset from Table~\ref{tab:data-basics} for a range of $\delta$ values. 
For the periodic datasets (i.e., \texttt{Hypertext}, \texttt{Infectious I}, and \texttt{Infectious II}), the $\delta$ values were chosen in line with the periodicity of measurements.
For the non-periodic datasets, $\delta$ values were chosen as a set number of hours, days, months, or even years.
For each dataset and $\delta$ value, we report in Tables~\ref{tab:temporal-results-small} and~\ref{tab:temporal-results-large} the number of $(\delta,1)$-maximal cliques found (that satisfy Definition~\ref{def:dgmax-clique}), the maximum cardinality (i.e., number of nodes in the largest clique) and maximum time interval duration among those cliques, and the runtime (in seconds) and maximum RAM usage (in MB) of each algorithm.
\afterpage{
\begin{landscape}
\begin{table}[t]
    \centering
    \caption{Results for the maximal $(\delta,1)$-clique count ($N$), maximum Cardinality ($C$), maximum clique duration, and the computation time (in seconds) and maximum space usage (in MB) for the small datasets.}
    \label{tab:temporal-results-small}
    \begin{footnotesize}  
    \begin{tabular}{ll|rrr|rr|rr|rr|rr}
         &  & \multicolumn{3}{c|}{Clique properties} & \multicolumn{2}{c|}{Viard et al.~\cite{viard2016computing}} & \multicolumn{2}{c|}{Himmel et al.~\cite{himmel2017adapting}} & \multicolumn{2}{c|}{Banerjee \& Pal~\cite{banerjee2024two}} & \multicolumn{2}{c}{Our alg. (BSN)} \\ \hhline{~~~~~--------}
        Dataset & $\delta$ & $N$ & $C$ & $D$ & \parbox{1.1cm}{\centering Runtime (\textit{s})} & \parbox{.9cm}{\centering Max. RAM (\textit{MB})} & \parbox{1.1cm}{\centering Runtime (\textit{s})} & \parbox{.9cm}{\centering Max. RAM (\textit{MB})} & \parbox{1.1cm}{\centering Runtime (\textit{s})} & \parbox{.9cm}{\centering Max. RAM (\textit{MB})} & \parbox{1.1cm}{\centering Runtime (\textit{s})} & \parbox{.9cm}{\centering Max. RAM (\textit{MB})} \\ \hline
        \multirow{10}{1.5cm}{Hypertext} 
            &  60 & 7,001 & 7 &  7,521 & 13.98 & 188 & 3.70 &$\leq$76 & 29.58 & 116 & \textbf{0.45} &$\leq$76 \\
            & 120 & 5,697 & 7 &  7,901 & 14.73 & 177 & 2.88 &$\leq$76 & 13.59 & 115 & \textbf{0.41} &$\leq$76 \\
            & 180 & 5,106 & 7 & 11,161 & 16.54 & 195 & 2.60 &$\leq$76 &  8.84 & 115 & \textbf{0.39} &$\leq$76 \\
            & 240 & 4,727 & 7 & 11,161 & 16.89 & 195 & 2.36 &$\leq$76 &  7.22 & 116 & \textbf{0.39} &$\leq$76 \\ 
            & 300 & 4,479 & 7 & 11,161 & 17.47 & 198 & 2.25 &$\leq$76 &  5.33 & 116 & \textbf{0.38} &$\leq$76 \\
            & 360 & 4,257 & 7 & 11,161 & 18.27 & 206 & 2.16 &$\leq$76 &  4.80 & 116 & \textbf{0.39} &$\leq$76 \\
            & 420 & 4,112 & 7 & 11,161 & 19.28 & 217 & 2.12 &$\leq$76 &  4.28 & 116 & \textbf{0.39} &$\leq$76 \\
            & 480 & 3,989 & 7 & 11,161 & 19.96 & 225 & 2.06 &$\leq$76 &  4.10 & 115 & \textbf{0.39} &$\leq$76 \\
            & 540 & 3,899 & 7 & 16,521 & 20.92 & 236 & 2.02 &$\leq$76 &  3.74 & 116 & \textbf{0.39} &$\leq$76 \\
            & 600 & 3,815 & 7 & 16,521 & 22.42 & 256 & 2.00 &$\leq$76 &  3.51 & 116 & \textbf{0.39} &$\leq$76 \\ \hline
        \multirow{5}{1.5cm}{College message} 
            &   3,600 & 33,350 & 4 &    14,562 &  25.82 & 249   & 25.07 &$\leq$76 & 11.07 & 150 & \textbf{0.61} &$\leq$76 \\
            &  43,200 & 23,713 & 5 &   316,619 &  31.82 & 422   & 19.50 &$\leq$76 & 2.64  & 145 & \textbf{0.54} &$\leq$76 \\
            &  88,640 & 19,925 & 5 &   718,855 &  38.69 & 605   & 17.92 &$\leq$76 & 1.57  & 143 & \textbf{0.54} &$\leq$76 \\
            & 259,200 & 16,262 & 5 & 1,804,213 &  60.71 & 1,163 & 16.66 &$\leq$76 & 1.05  & 143 & \textbf{0.56} &$\leq$76 \\
            & 604,800 & 14,583 & 6 & 5,124,654 & 103.34 & 2,342 & 16.17 &$\leq$76 & 0.93  & 144 & \textbf{0.66} &$\leq$76 \\ \hline
        \multirow{5}{1.5cm}{Bitcoin} 
            &   3,600 & 25,541 & 7 &   3,592 & 13.82 & 111 & 124.77 &$\leq$76 & 1.88 & 158 & \textbf{0.60} &$\leq$76 \\
            &  43,200 & 24,178 & 8 &  47,301 & 14.85 & 123 & 125.00 &$\leq$76 & 1.86 & 158 & \textbf{0.64} &$\leq$76 \\
            &  88,640 & 23,580 & 8 & 168,288 & 15.57 & 138 & 124.84 &$\leq$76 & 1.84 & 159 & \textbf{0.67} &$\leq$76 \\
            & 259,200 & 22,558 & 8 & 259,173 & 17.37 & 174 & 125.49 &$\leq$76 & 1.83 & 160 & \textbf{0.71} &$\leq$76 \\
            & 604,800 & 21,608 & 8 & 610,794 & 22.16 & 252 & 125.34 &$\leq$76 & 1.92 & 162 & \textbf{0.80} &$\leq$76 \\ \hline
        \multirow{6}{1.5cm}{Infectious I} 
            &     60 & 115,612 &  6 &  3,641 &   168.67 & 2,991   & 466.49 & 156 & 27.32 & 329   &  \textbf{6.36} & 229  \\ 
            &    120 &  88,394 &  6 &  4,941 &   252.29 & 4,291   & 461.02 & 143 & 15.83 & 328   &  \textbf{6.86} & 210  \\ 
            &    600 &  68,807 & 10 & 10,001 & 1,618.69 & 21,795  & 465.47 & 140 & \textbf{12.87} & 481   & 15.42 & 189 \\ 
            &  1,200 &  61,530 & 13 & 10,001 & 8,493.14 & 90,185  & 485.72 & 152 & \textbf{30.63} & 923   & 39.05 & 183 \\ 
            &  6,000 &  21,720 & 16 & 10,741 & \multicolumn{2}{c|}{$>200$\textit{GB}} & 581.36 & 165 & 63.38 & 2,846 & \textbf{14.94} & 139 \\ 
            & 12,000 &  20,749 & 16 & 11,301 & \multicolumn{2}{c|}{$>200$\textit{GB}} & 580.86 & 165 & 63.08 & 2,864 &  \textbf{5.21} & 137  \\ \hline
        \multirow{10}{1.5cm}{Infectious II} 
            & 60  & 6,446 & 5 & 1,741 &   5.80 & 114   & 2.12 &$\leq$76 & 1.88 & 117 & \textbf{0.35} &$\leq$76 \\
            & 120 & 5,358 & 6 & 2,661 &  10.08 & 186   & 1.88 &$\leq$76 & 0.96 & 118 & \textbf{0.42} &$\leq$76 \\
            & 180 & 5,227 & 6 & 3,921 &  14.94 & 271   & 1.79 &$\leq$76 & 0.73 & 119 & \textbf{0.50} &$\leq$76 \\
            & 240 & 5,240 & 7 & 4,681 &  21.76 & 372   & 1.79 &$\leq$76 & 0.68 & 121 & \textbf{0.58} &$\leq$76 \\
            & 300 & 5,334 & 8 & 4,681 &  29.93 & 487   & 1.84 &$\leq$76 & 0.72 & 122 & \textbf{0.67} &$\leq$76 \\
            & 360 & 5,512 & 8 & 5,761 &  40.46 & 625   & 1.94 &$\leq$76 & \textbf{0.78} & 126 & 0.81 &$\leq$76 \\
            & 420 & 5,835 & 8 & 5,761 &  55.27 & 825   & 2.09 &$\leq$76 & \textbf{0.85} & 131 & 0.95 &$\leq$76 \\
            & 480 & 6,004 & 9 & 7,581 &  71.67 & 1,034 & 2.25 &$\leq$76 & \textbf{1.04} & 134 & 1.13 &$\leq$76 \\
            & 540 & 6,133 & 9 & 8,501 &  94.75 & 1,302 & 2.42 &$\leq$76 & \textbf{1.08} & 137 & 1.30 &$\leq$76 \\
            & 600 & 6,266 & 9 & 8,501 & 121.87 & 1,625 & 2.60 &$\leq$76 & \textbf{1.19} & 140 & 1.50 &$\leq$76 \\ \hline
    \end{tabular}
    \end{footnotesize}
\end{table}

\begin{table}[t]
    \centering
    \caption{Results for the maximal $(\delta,1)$-clique count ($N$), maximum Cardinality ($C$), maximum clique duration, and the computation time (in seconds) and maximum space usage (in MB) for the larger datasets.}
    \label{tab:temporal-results-large}
    \begin{footnotesize}  
    \begin{tabular}{ll|rrr|rr|rr|rr|rr}
         &  & \multicolumn{3}{c|}{Clique properties} & \multicolumn{2}{c|}{Viard et al.~\cite{viard2016computing}} & \multicolumn{2}{c|}{Himmel et al.~\cite{himmel2017adapting}} & \multicolumn{2}{c|}{Banerjee \& Pal~\cite{banerjee2024two}} & \multicolumn{2}{c}{Our alg. (BSN)} \\ \hhline{~~~~~--------}
        Dataset & $\delta$ & $N$ & $C$ & $D$ & \parbox{1.1cm}{\centering Runtime (\textit{s})} & \parbox{.9cm}{\centering Max. RAM (\textit{MB})} & \parbox{1.1cm}{\centering Runtime (\textit{s})} & \parbox{.9cm}{\centering Max. RAM (\textit{MB})} & \parbox{1.1cm}{\centering Runtime (\textit{s})} & \parbox{.9cm}{\centering Max. RAM (\textit{MB})} & \parbox{1.1cm}{\centering Runtime (\textit{s})} & \parbox{.9cm}{\centering Max. RAM (\textit{MB})} \\ \hline
        \multirow{6}{1.25cm}{Facebook-wosn-wall} 
            &      86,400 & 506,882 &  4 &   3,373,951 &   305.85 & 2,547   & 10,286.43 & 532 & 562.87 & 771 & \textbf{11.97} & 619 \\
            &     604,800 & 380,549 &  5 &  21,687,693 &   397.67 & 4,488   & 10,354.96 & 469 &  63.70 & 704 & \textbf{11.27} & 553 \\
            &   2,628,000 & 281,712 &  6 &  79,749,744 &   757.84 & 11,377  & 10,492.18 & 425 &  20.31 & 661 & \textbf{11.52} & 493 \\
            &   7,884,000 & 218,340 &  7 & 118,451,848 & 1,724.46 & 30,867  & 10,539.35 & 406 &  14.79 & 660 & \textbf{12.41} & 470 \\
            &  31,536,000 & 170,460 &  9 & 129,313,589 & 8,333.04 & 200,573 & 10,287.60 & 381 &  \textbf{14.00} & 691 & 15.08 & 459 \\
            & 157,680,000 & 131,066 & 10 & 131,845,082 & \multicolumn{2}{c|}{$>200$\textit{GB}} & 10,375.97 & 380 &  14.68 & 744 & \textbf{11.63} & 433 \\ \hline
        \multirow{3}{1.25cm}{Reddit hyperlinks} 
            &   3,600     & 839,208 &  5 &      5,190 &  3,952.44 &   1,988 & 28,488.17 & 825 & \multicolumn{2}{c|}{$>24$h} & \textbf{140.38} & 4.765 \\
            &  86,400     & 789,675 &  9 &    255,291 &  4,326.71 &   3,582 & 28,385.90 & 833 & \multicolumn{2}{c|}{$>24$h} & \textbf{151.55} & 4.923 \\
            & 604,800     & 701,663 & 17 & 73,872,286 & 31,687.53 & 127,349 & 27,969.37 & 886 & 21,697.68 & 12,109 & \textbf{628.46} & 5.143 \\ \hline
        \multirow{4}{1.25cm}{Enron} 
            &     3,600   & 983,610 & 12 &     29,588 & 2,761.12 & 21,014 & 43,035.89 &   966 & \multicolumn{2}{c|}{$>24$h} &    \textbf{80.17} & 2.664 \\
            &    86,400   & 721,768 & 13 &  1,054,487 & 8,489.00 & 73,895 & 41,450.11 &   926 & 14,019.07 &  2,439 &   \textbf{107.12} & 2.288 \\
            &   604,800   & 591,242 & 15 & 39,517,939 & \multicolumn{2}{c|}{$>200$\textit{GB}} & 39,910.54 &   895 & 55,401.08 &  5,981 &   \textbf{286.77} & 1.899 \\
            & 2,628,000   & 599,179 & 18 & 66,570,802 & \multicolumn{2}{c|}{$>200$\textit{GB}} & 42,054.39 & 1,009 & 13,654.13 & 49,110 & \textbf{2,634.49} & 1.732 \\ \hline
        \multirow{4}{1.25cm}{Last.fm bands} 
            &      86,400 &  5,937,774 &  5 &   7,213,683 & \multicolumn{2}{c|}{$>24$h} & \multicolumn{2}{c|}{$>24$h} & \multicolumn{2}{c|}{$>24$h} &    \textbf{897.09} & 26,059 \\
            &     604,800 &  4,101,745 &  7 &  79,771,958 & \multicolumn{2}{c|}{$>24$h} & \multicolumn{2}{c|}{$>24$h} & 63,689.40 &  10,251 &    \textbf{944.14} & 20,509 \\
            &   2,628,000 &  4,899,608 & 10 & 134,178,398 & \multicolumn{2}{c|}{$>24$h} & \multicolumn{2}{c|}{$>24$h} & 13,433.54 &  24,517 &  \textbf{1,904.25} & 22,541 \\
            &   7,884,000 & 12,153,968 & 14 & 136,945,742 & \multicolumn{2}{c|}{$>24$h} & \multicolumn{2}{c|}{$>24$h} & 18,478.72 & 135,215 & \textbf{10,458.64} & 43,456 \\ \hline
        \multirow{5}{1.25cm}{Last.fm songs} 
            &      86,400 & 14,715,165 &  4 &   3,342,331 & \multicolumn{2}{c|}{$>24$h} & \multicolumn{2}{c|}{$>24$h} & \multicolumn{2}{c|}{$>24$h} &   \textbf{575.71} & 20,637 \\
            &     604,800 & 11,552,954 &  5 &  24,493,086 & \multicolumn{2}{c|}{$>24$h} & \multicolumn{2}{c|}{$>24$h} & 56,838.23 & 16,692 &   \textbf{536.07} & 17,565 \\
            &   2,628,000 &  8,647,218 &  6 & 120,376,971 & \multicolumn{2}{c|}{$>24$h} & \multicolumn{2}{c|}{$>24$h} & 35,852.11 & 15,040 &   \textbf{675.74} & 14,769 \\
            &   7,884,000 &  7,429,520 &  8 & 134,177,875 & \multicolumn{2}{c|}{$>24$h} & \multicolumn{2}{c|}{$>24$h} & 27,454.15 & 16,404 & \textbf{1,044.68} & 14,037 \\
            &  31,536,000 &  9,853,895 & 11 & 136,828,008 & \multicolumn{2}{c|}{$>24$h} & \multicolumn{2}{c|}{$>24$h} & 22,874.81 & 38,107 & \textbf{3,186.39} & 17,576 \\ 
            & 157,680,000 & 10,491,336 & 17 & 156,295,069 & \multicolumn{2}{c|}{$>24$h} & \multicolumn{2}{c|}{$>24$h} & \multicolumn{2}{c|}{$>200$\textit{GB}} & \textbf{6,078.19} & 17,644 \\ \hline
    \end{tabular}
    \end{footnotesize}
\end{table}
\end{landscape}
}

\begin{figure}[t]
    \centering
    \includegraphics[width=\linewidth]{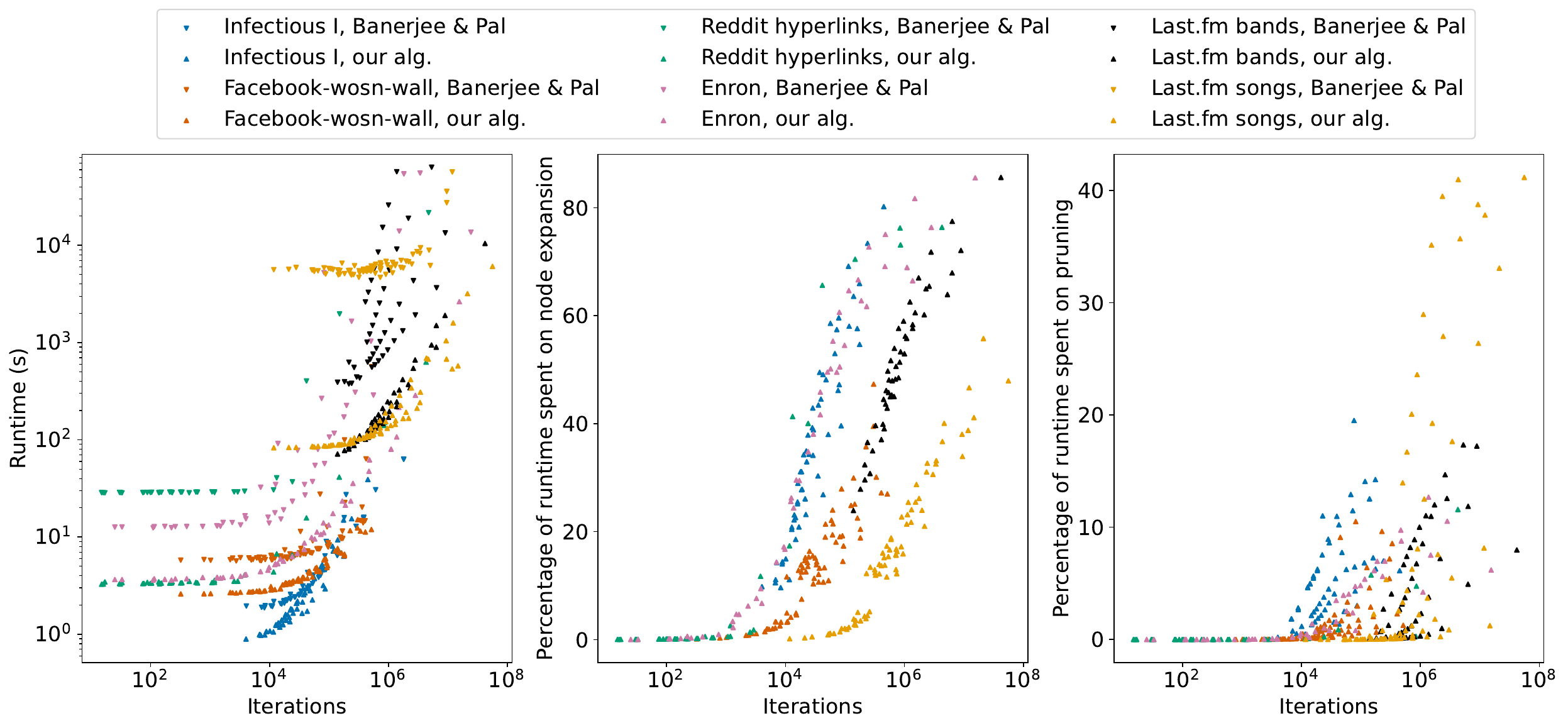}
    \caption{Runtime vs. the number of iterations (search space states) that are processed (left panel). The centre and right panel show respectively the share of runtime of our proposed algorithm used on node expansion and pruning decision making with respect to the number of iterations. Results for all $\delta$ and $\gamma$ values tested for the given datasets are included, except when they resulted in zero ($\delta,\gamma$)-maximal cliques.}
    \label{fig:iter-time}
\end{figure}

Results from Tables~\ref{tab:temporal-results-small} and~\ref{tab:temporal-results-large} show that for all but a few instances, our proposed algorithm is faster than existing methods.
Especially for larger datasets and at relatively low $\delta$ values, we significantly outperform Banerjee \& Pal.
Note that, for the instances where we do not improve upon them, we still required fewer iterations, i.e., processed search tree states (see the left panel of Figure~\ref{fig:iter-time}).
However, depending on the specific scenario, computing the time interval borders of our corrected definition of $(\delta,\gamma)$-cliques (as described in Section~\ref{subsubsect:methods-overlap}), may lead to computationally heavier iterations than for Banerjee \& Pal. 
Thus, although rare, when insufficient pruning occurs to offset this, our algorithm can under perform.

Figure~\ref{fig:iter-time} shows the relationship between the number of iterations (i.e., search tree states) that must be processed and the runtime. 
For each dataset we observe that, as the number of required iterations increases, the relative amount of time spent on node expansion computation increases as well.
Furthermore, the time spent on node expansion tends to far exceed that spent on the pruning decision making, thus highlighting its relative efficiency.
As such, two factors appear to have the most significant impact on the performance of our algorithm.
First, the maximum cardinality, which can drastically increase the number of iterations; and second, the effect of the chosen $\delta$ on the extent to which we may prune the search space.
Results from Tables~\ref{tab:temporal-results-small} and~\ref{tab:temporal-results-large} suggest that, when different $\delta$ lead to similar maximum cardinality, performance is lowest for mid-size $\delta$.
For smaller $\delta$ there are relatively smaller cliques in general, but there is also less time interval overlap to account for, leading to more pruning.
Meanwhile, for larger $\delta$ that approach (or exceed) the full lifetime of the dataset, temporal growth dominance becomes a given, which aids pruning.

In terms of space (i.e., RAM) requirements, we observe that our algorithm generally requires less memory than Banerjee \& Pal. 
Only on larger datasets with many $(\delta,\gamma)$-maximal cliques, do we sometimes require more space due to the additional clique properties that are stored.
This could easily be compensated for by writing the $(\delta,\gamma)$-maximal cliques to a file as soon as they are found, instead of storing them in a set during enumeration.
After all, unlike existing methods we prevent duplicates through our search order, removing the necessity of checks for duplicates.
However, we did not implement this in order to maintain a fair comparison with existing methods for our experiments and to allow for easier integration with future coding.

Compared to Himmel et al., we require more space due to the use of additional data structures.
However, Table~\ref{tab:temporal-results-large} shows that the approach by Himmel et al. requires more computation time and scales poorly to larger datasets.
Finally, Viard et al. shows worse performance in both runtime and space requirements, with space requirements exploding for larger $\delta$.

\subsubsection{Comparison for larger \texorpdfstring{$\gamma$}{gamma} in the unweighted setting}
\label{subsubsect:results-unw}
\afterpage{
\begin{figure}[t]
    \centering
    \begin{subfigure}[t]{\linewidth}
        \centering
        \includegraphics[width=\linewidth]{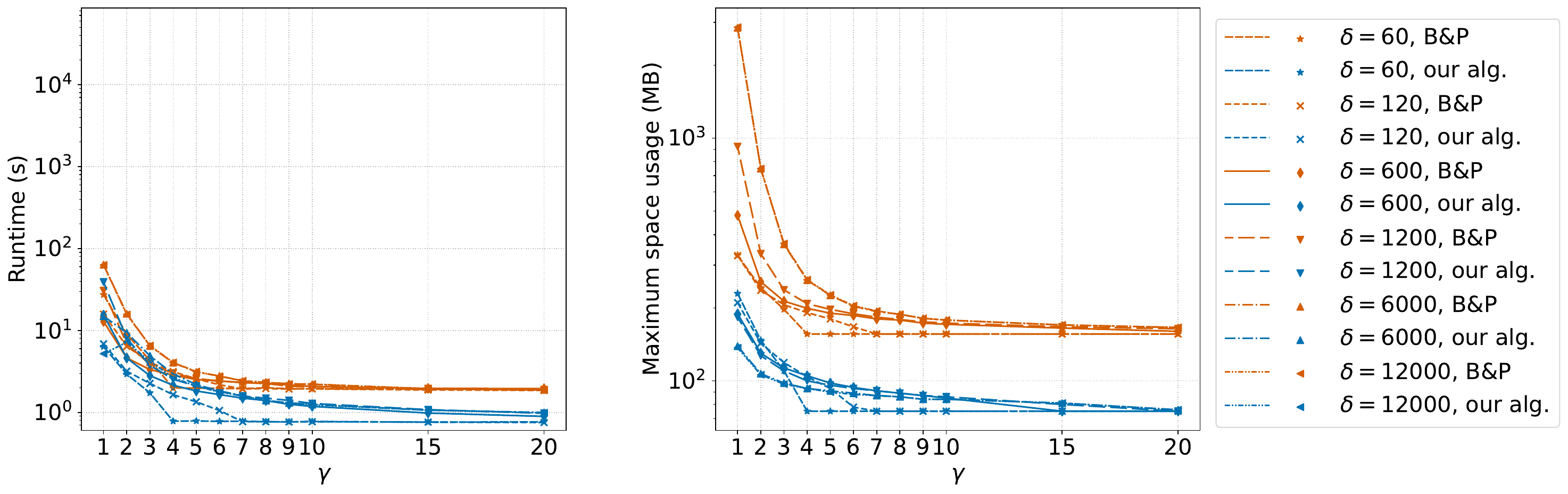}
        \caption{Infectious I}
    \end{subfigure}
    \begin{subfigure}[t]{\linewidth}
        \centering
        \includegraphics[width=\linewidth]{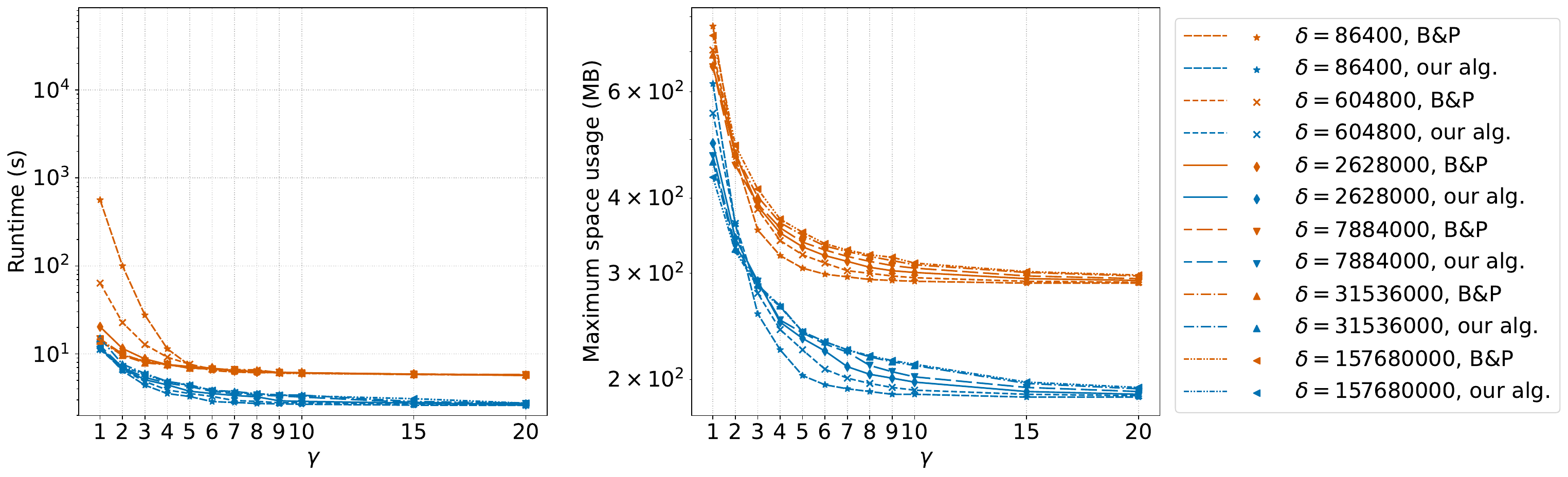}
        \caption{Facebook-wosn-wall}
    \end{subfigure}
    \begin{subfigure}[t]{\linewidth}
        \centering
        \includegraphics[width=\linewidth]{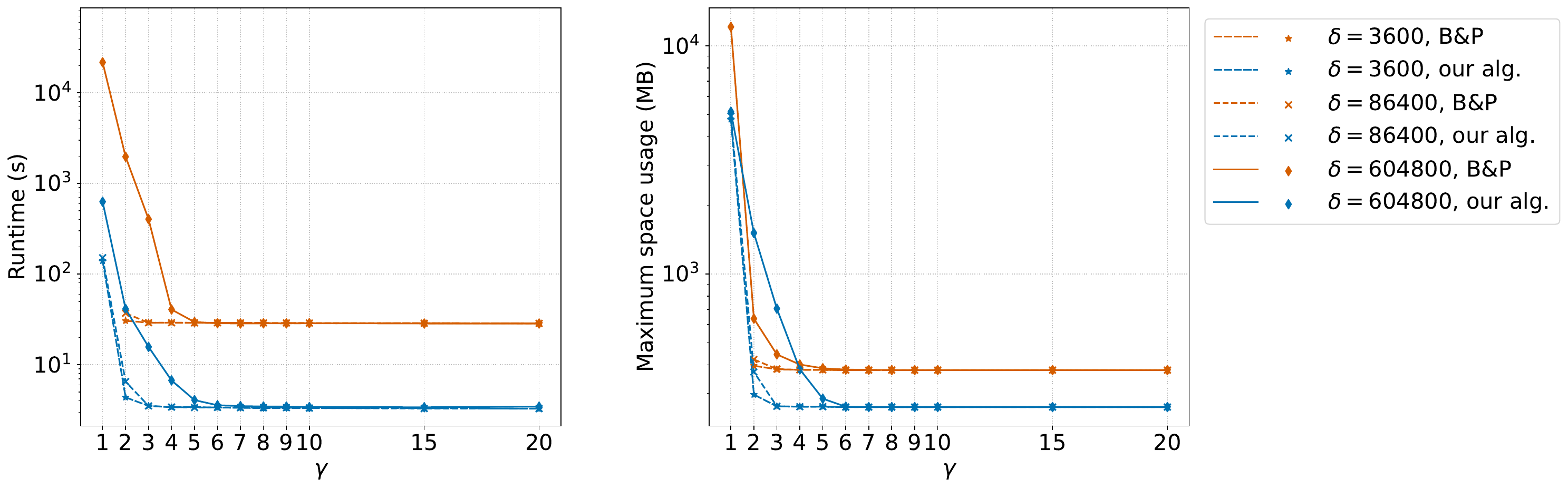}
        \caption{Reddit Hyperlinks}
    \end{subfigure}
    \caption{Performance comparison between Banerjee \& Pal~\cite{banerjee2024two} and our algorithm for increasing $\gamma$. Results for $\gamma=1$ correspond to those reported in Tables~\ref{tab:temporal-results-small} and~\ref{tab:temporal-results-large}.}
    \label{fig:res-gammacomp-performance-1}
\end{figure}
\begin{figure}[t]
    \centering
    \begin{subfigure}[t]{\linewidth}
        \centering
        \includegraphics[width=\linewidth]{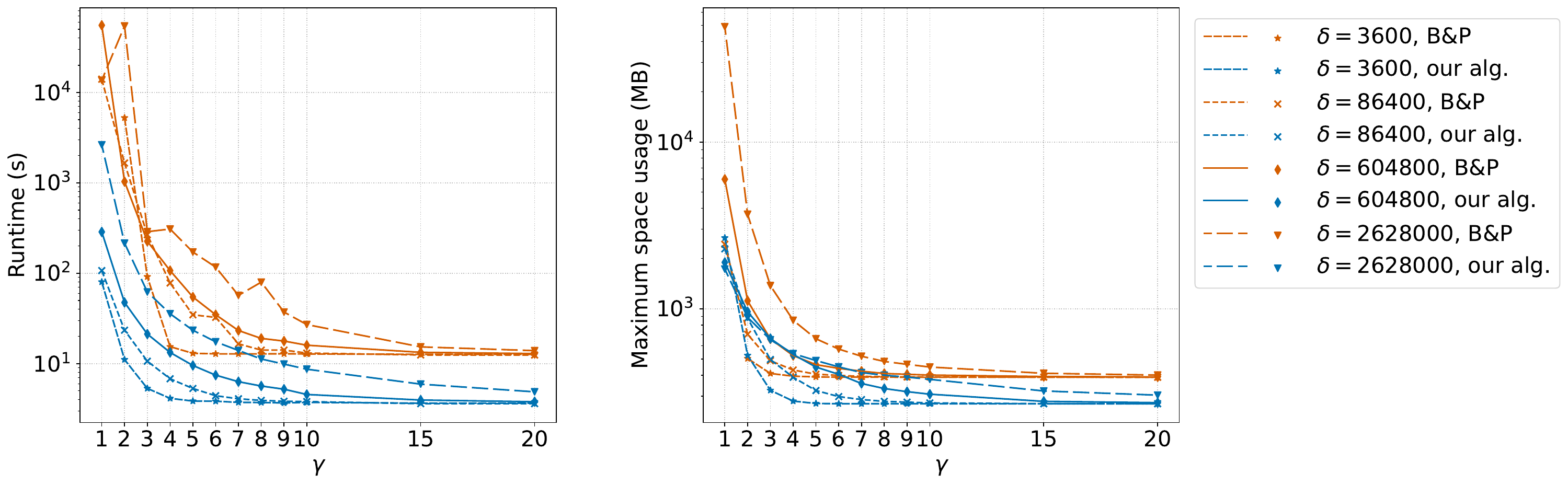}
        \caption{Enron}
    \end{subfigure}  
    \begin{subfigure}[t]{\linewidth}
        \centering
        \includegraphics[width=\linewidth]{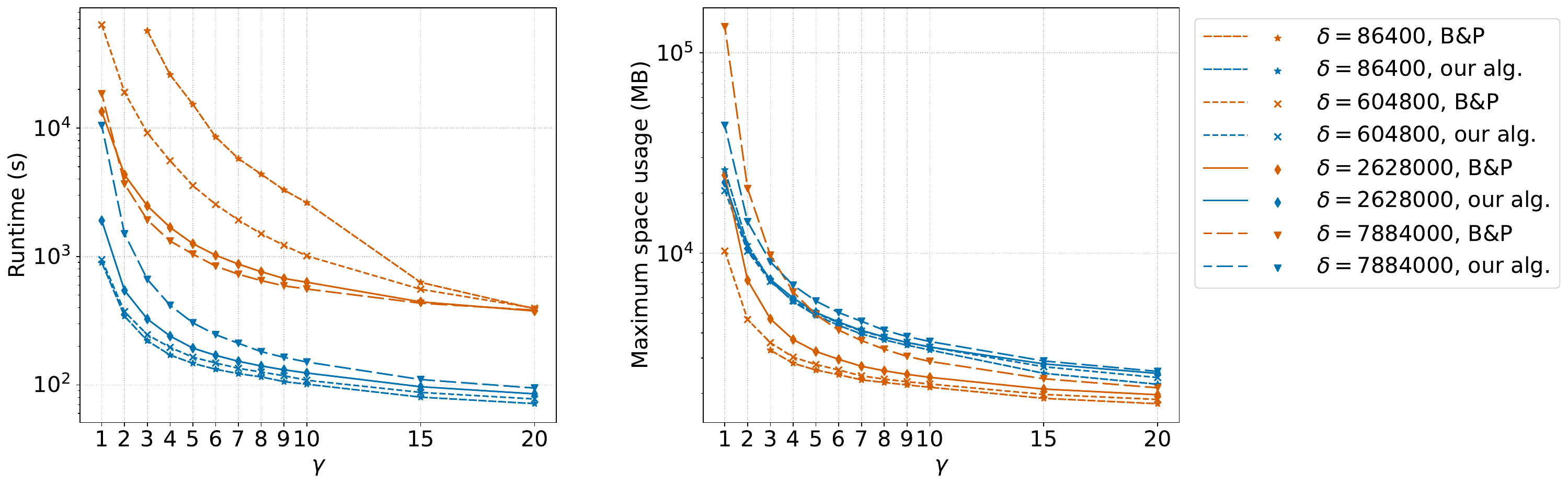}
        \caption{Last.fm bands}
    \end{subfigure} 
    \begin{subfigure}[t]{\linewidth}
        \centering
        \includegraphics[width=\linewidth]{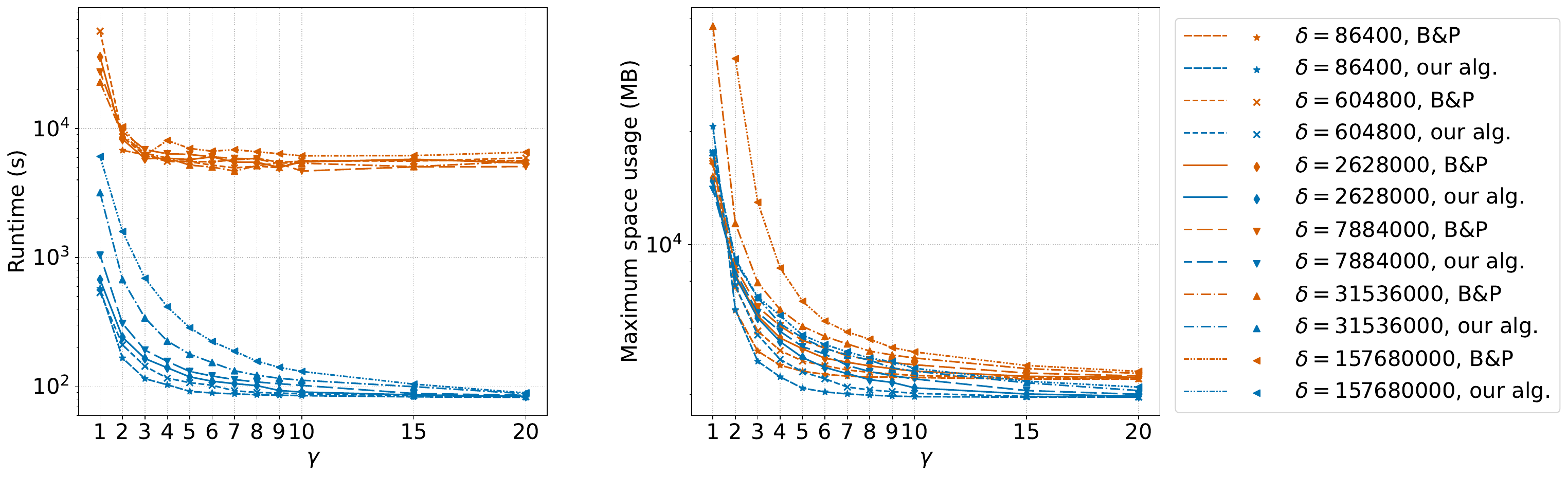}
        \caption{Last.fm songs}
    \end{subfigure}  
    \caption{Performance comparison between Banerjee \& Pal~\cite{banerjee2024two} and our algorithm for increasing $\gamma$. Results for $\gamma=1$ correspond to those reported in Tables~\ref{tab:temporal-results-small} and~\ref{tab:temporal-results-large}.}
    \label{fig:res-gammacomp-performance-2}
\end{figure}
}
Having demonstrated that our algorithm outperforms existing algorithms in (nearly) all instances in the basic temporal setting, we next compare against Banerjee \& Pal's algorithm for a range of $\gamma$ values in the unweighted setting.
Performance results for the six largest datasets are shown in Figures~\ref{fig:res-gammacomp-performance-1} and~\ref{fig:res-gammacomp-performance-2}.
Information on the $(\delta,\gamma)$-clique properties are provided in Figures~\ref{fig:res-gammacomp-properties-1} and~\ref{fig:res-gammacomp-properties-2} in the Appendix.

The results clearly demonstrate that, as $\gamma$ increases, the number of cliques and consequently the runtime and memory requirements decrease.
Furthermore, they demonstrate that our algorithm outperforms Banerjee \& Pal both computationally and in space requirements.
Moreover, for the instances where Banerjee \& Pal performed better at $\gamma = 1$, our algorithm performs better at higher $\gamma$ values.

\subsubsection{Comparing the weighted and unweighted settings}
\label{subsubsect:results-weighted}
To conclude our experiments, we compared the weighted and unweighted settings for the two weighted datasets (\texttt{Bitcoin} and \texttt{Reddit hyperlinks}), given identical $\delta$ and $\gamma$ values.
The results, shown and analyzed in Appendix~\ref{apsect:results-weighted}, demonstrate that the weighted setting allows better fine-tuning of the (positive) strength of connections beyond mere frequency.
In other words, the weighted setting allows us to consider the sentiment of a connection when forming cliques.
Furthermore, results confirmed that $\gamma$ itself (whether weighted or unweighted) has no direct impact on the runtime or space complexities of our algorithm. However, due to its influence on the number of (valid) $(\delta,\gamma)$-cliques, it does impact the performance indirectly.

\section{Conclusions}
\label{sect:conclusions}
In this work, we introduced a $(\delta,\gamma)$-maximal clique enumeration algorithm that improves upon the state-of-the-art two-phase approach introduced by Banerjee \& Pal~\cite{banerjee2024two}.
Furthermore, we reformulated the definition of $(\delta,\gamma)$-cliques to fix two issues in the temporal domain and to follow a more natural interpretation of $\delta$. 
Additionally, we extended the definition to cover weighted networks.

Our proposed algorithm improves the time complexity of the initial stretching phase such that it is linear with respect to the number of temporal edges, while also accounting for edge weighting.
Although our corrected definition requires more complex considerations, during clique node set expansions, in the subsequent bulking phase, we improved its computation time by introducing an efficient pruning method.
Furthermore, by employing a custom node labeling and associated search order, we were able to reduce the space requirements and skip any duplicate search states.

Our experiments empirically demonstrated a substantial performance improvement by our algorithm in both computational time and space requirements over existing methods.
Furthermore, we showed that by extending the methodology to be able to handle weighted networks, we have opened up more nuanced options for imposing minimum requirements on clique formation.
Thus, in summary, the method we introduced in this paper is easier to interpret, scales better to larger networks, and allows one to impose more real-world restrictions on cliques through weighted networks.

Given that computational performance is tied to the number of search states processed, future improvements on this method could consider other approaches to node labeling that (further) optimizes the amount of pruning that is achieved.
Such methods could try to combine pivot selections along the lines of those used for static cliques, while also taking into account the temporal dominance.
Another potential improvement can be made by parallelizing the two phases of the algorithm. After all, in the stretching phase each edge can be processed independently, while in the bulking phase processing each clique in $C^s$ independently may at worst lead us to process cliques whose search tree branches could have been pruned.

\section*{Statements and Declarations}
\subsection*{Author contributions}
Conceptualization: Hanjo Boekhout; Methodology: Hanjo Boekhout; Implementation and experimental analysis: Hanjo Boekhout; Writing - original draft preparation: Hanjo Boekhout; Writing - review and editing: Frank Takes and Hanjo Boekhout; Supervision: Frank Takes.

\subsection*{Funding}
No funding was received for conducting this study.

\subsection*{Competing interests}
The authors have no competing interests to declare that are relevant to the content of this article.

\subsection*{Code availability}
The (modified) implementations of all algorithms used in the experiments of this study can be found at \url{https://github.com/hdboekhout/Fast-delta-gamma-clique}.

\subsection*{Data availability}
All datasets used in this work were obtained from online accessible data repositories. Appropriate citation/reference linkages have been included for each dataset (see Table~\ref{tab:data-basics}).

\bibliography{bibliography}

\clearpage
\appendix
\section{Algorithm pseudo code} \label{app:A}
\begin{algorithm}
    \caption{Stretch phase wrapper function}\label{alg:stretch-wrapper}
    \begin{scriptsize}
    \begin{algorithmic}[1]
        \Require Given (weighted) temporal graph $\mathcal{G} = (V, E, \mathcal{T}, \mathcal{W^T})$, let $\mathit{all\_times}$ and $\mathit{all\_weights}$ be dictionaries of lists such that for each edge $(t_i,u_i,v_i,w_i) \in E^{\mathcal{TW^T}}$, with $o_i$ indicating its place in the temporally ordered sequence of edge instances of $(u_i,v_i)$, it holds that $\mathit{all\_times}[(u_i,v_i)][o_i] = t_i$ and $\mathit{all\_weights}[(u_i,v_i)][o_i] = w_i$.
        \Ensure $C^s$ is the set of all 2-node ($\delta,\gamma$)-cliques that are duration-wise maximal
        \Statex
        \Procedure{stretchPhase}{$\delta, \gamma$}
            \If{$\delta > 0$}
                \For{$(u,v) \in E$} \label{line:stretch-start}
                    \If{sum($\mathit{all\_weights}[(u,v)]) \geq \gamma$} \label{line:stretch-start-weightcheck}
                        \State \Call{Stretch}{$\{u,v\}, \delta, \gamma, \mathit{all\_times}[(u,v)], \mathit{all\_weights}[(u,v)]$}
                    \EndIf
                \EndFor
            \EndIf
        \EndProcedure
    
    \end{algorithmic}
    \end{scriptsize}
\end{algorithm}

\begin{algorithm}
    \caption{Stretch phase function for a given link}\label{alg:stretch-edge}
    \begin{scriptsize}
    \begin{algorithmic}[1]
        \setcounter{ALG@line}{9}
        \Procedure{Stretch}{$\mathit{node\_set} = \{u,v\}, \delta, \gamma, \mathit{times}, \mathit{weights}$} \label{line:stretch-edge-start}
            \State $\mathit{bi}, \mathit{ei} \gets 0, 0$  \label{line:stretch-11}
            \State $\mathit{current\_weight} \gets \mathit{weights}[\mathit{bi}]$ \label{line:stretch-12}

            \While{$True$}
                \While{$\mathit{current\_weight} < \gamma$} \Comment{Find next set of edges with sufficient weight}  \label{line:stretch-14}
                    \State $\mathit{ei} \gets \mathit{ei} + 1$
                    \If{$\mathit{ei} == len(\mathit{times})$} \Return  \label{line:stretch-16}
                    \EndIf                                            
                    \While{$\mathit{times}[\mathit{ei}] - \mathit{times}[\mathit{bi}] \geq \delta$}  \label{line:stretch-18}
                        \State $\mathit{current\_weight} \gets \mathit{current\_weight} - \mathit{weight}[\mathit{bi}]$
                        \State $\mathit{bi} \gets \mathit{bi} + 1$
                    \EndWhile    \label{line:stretch-21}
                    \State $\mathit{current\_weight} \gets \mathit{current\_weight} + \mathit{weight}[\mathit{ei}]$ 
                \EndWhile 
                \State $\mathit{tb}, \mathit{tbMax} \gets \mathit{times}[\mathit{bi}], \mathit{times}[\mathit{ei}]$  \label{line:stretch-24}
                \Comment{Remember clique start borders}
                \Statex
                \State $\mathit{done} = False$ \label{line:stretch-25}
                \While{not $\mathit{done}$}
                    \State $\mathit{old\_bi}, \mathit{old\_ei} \gets \mathit{bi}, \mathit{ei}$
                    \While{$\mathit{ei} + 1 < len(\mathit{times})$ and $\mathit{times}[\mathit{ei} + 1] - \mathit{times}[bi] < \delta$} \label{line:stretch-28}
                        \State $\mathit{ei} \gets \mathit{ei} + 1$
                        \State $\mathit{current\_weight} \gets \mathit{current\_weight} + \mathit{weight}[\mathit{ei}]$ 
                    \EndWhile \label{line:stretch-31}  
                    \While{$\mathit{bi} + 1 \leq \mathit{ei}$ and $\mathit{current\_weight} - \mathit{weights}[bi] \leq \gamma$} \label{line:stretch-32}
                        \State $\mathit{current\_weight} \gets \mathit{current\_weight} - \mathit{weight}[\mathit{bi}]$
                        \State $\mathit{bi} \gets \mathit{bi} - 1$
                    \EndWhile \label{line:stretch-35}  
                    \If{$\mathit{ei} + 1 == len(\mathit{times})$} \label{line:stretch-36}
                        \State $C^s$.addClique($\mathit{node\_set}, (\mathit{tb}, \mathit{tbMax}, \mathit{times}[\mathit{bi}], \mathit{times}[\mathit{ei}])$)
                        \State \Return
                    \EndIf \label{line:stretch-39}
                    \Statex
                    \If{$\mathit{bi} == \mathit{old\_bi}$ and $\mathit{ei} == \mathit{old\_ei}$} 
                        \State $\mathit{bt}, \mathit{et} \gets \mathit{times}[\mathit{bi}] + 1, \mathit{ei}$ \label{line:stretch-41}
                        \State $\mathit{current\_weight} \gets \mathit{current\_weight} - \mathit{weight}[\mathit{bi}]$
                        \While{$\mathit{current\_weight} < \gamma$}
                            \State $\mathit{et} \gets \mathit{et} + 1$
                            \If{$\mathit{et} == len(\mathit{times})$}  \label{line:stretch-45}
                                \State $C^s$.addClique($\mathit{node\_set}, (\mathit{tb}, \mathit{tbMax}, \mathit{times}[\mathit{bi}], \mathit{times}[\mathit{ei}])$)
                                \State \Return 
                            \EndIf \label{line:stretch-48}
                            \If{$\mathit{times}[\mathit{et}] - \mathit{bt} \geq \delta$} \label{line:stretch-49}
                                \State $C^s$.addClique($\mathit{node\_set}, (\mathit{tb}, \mathit{tbMax}, \mathit{times}[\mathit{bi}], \mathit{times}[\mathit{ei}])$)
                                \State $\mathit{done} \gets True$
                                \State $\mathit{et} \gets \mathit{et} - 1$ \Comment{Reduce the index as its weight was not yet added}
                                \State \textbf{break}
                            \EndIf \label{line:stretch-54}
                            \State $\mathit{current\_weight} \gets \mathit{current\_weight} + \mathit{weight}[\mathit{et}]$                     
                        \EndWhile
                        \State $\mathit{bi} \gets \mathit{bi} + 1$
                        \State $\mathit{ei} \gets \mathit{et}$ \label{line:stretch-58}
                    \EndIf
                \EndWhile    \label{line:stretch-60}
            \EndWhile       
        \EndProcedure \label{line:stretch-edge-end}
    \end{algorithmic}
    \end{scriptsize}
\end{algorithm}

\begin{algorithm}
    \caption{Bulk phase wrapper and helper functions}\label{alg:bulk-wrapper}
    \begin{scriptsize}
    \begin{algorithmic}[1]
        \setcounter{ALG@line}{62}
        \Require $C^s$ is the set of all 2-node duration-wise maximal ($\delta,\gamma$)-cliques defined as triplets $C = (X, [t_b, t_e], (tbMin, tbMax, teMin, teMax))$; $\cdots$
        \Ensure $R$ is the set of all $(\delta,\gamma)$-maximal cliques
        \Statex
        \Procedure{bulkPhase}{$\delta$}
            \State $D \gets \emptyset$
            \For{$C \in C^s$} \Comment{Cliques processed in increasing order based on $\max(X)$}
                \If{$c \notin D$}
                    \State $D \gets D \cup \Call{bulkRecursiveOuter}{\delta, C}$
                \EndIf
            \EndFor
        \EndProcedure
        \Statex
        \Procedure{isTemporallyDominated}{$C, C_{dom}$} \label{line:tempdom-start}
            \If{$C_{dom}.t_b \leq C.t_b$ and $C.t_e \leq C_{dom}.t_e$}
                \State \Return True      \Comment{Time interval of $C$ is fully covered by $C_{dom}$}
            \EndIf
            \State \Return False
        \EndProcedure  \label{line:tempdom-end}
        \Statex
        \Procedure{isSpatialGrowthDominated}{$C, N^e, C_{new}, E^r$} \label{line:spatial-growth-start}
            \If{$\{u \in N(C.X) \setminus N^e | u > \max(C.X)\} = \emptyset$}
                \If{$\max(E^r.\mathit{tbMin}) \leq C_{new}.\mathit{teMin}$ and $C_{new}.\mathit{tbMax} \leq \min(E^r.\mathit{teMax}) $}
                    \State \Return True      \Comment{All potential extensions to $C.X$ are covered by $N^e$}
                \EndIf
            \EndIf
            \State \Return False
        \EndProcedure \label{line:spatial-growth-end}
        \Statex
        \Procedure{isTemporalGrowthDominated}{$C, E^r, N^r, C^r, C^p$} \label{line:temporal-growth-start}
            \If{$\max(E^r.\mathit{tbMin}) \leq C.\mathit{tbMin}$ and $C.\mathit{teMax} \leq \min(E^r.\mathit{teMax})$} \label{line:cut-tg-1-start}
                \State \Return True  \Comment{Minimum potential growth $E^r$ exceeds potential growth of $C$}
            \EndIf \label{line:cut-tg-1-end}
            \Statex
            \If{$\min(C^r.t_b) < \max(E^r.\mathit{tbMin})$ or $\min(E^r.\mathit{teMax}) < \max(C^r.t_e)$} \label{line:cut-tg-2-start}
                \State \Return False   \Comment{Original time intervals exceed minimum potential growth of $E^r$}
            \EndIf \label{line:cut-tg-2-end}
            \Statex
            \For{$C_{i} \in \{C_j \in C^s | C_j.X \subseteq N^r\}$}  \Comment{Check edges between not yet added neighbors} \label{line:cut-tg-3-start}
                \If{$C_i.\mathit{teMax} \geq \min(C^p.\mathit{tbMax})$ and $C_i.t_b < \max(E^r.\mathit{tbMin})$}
                    \State \Return False   \Comment{Sufficient overlap with $C$, yet before potential growth of $E^r$}
                \EndIf
                \If{$C_i.\mathit{tbMin} \leq \max(C^p.\mathit{teMin})$ and $C_i.t_e > \min(E^r.\mathit{teMax})$}
                    \State \Return False \Comment{Sufficient overlap with $C$, yet after potential growth of $E^r$}
                \EndIf
            \EndFor \label{line:cut-tg-3-end}
            \State \Return True \Comment{No actual growth beyond potential of $E^r$ possible for $C$, safe to cut}
        \EndProcedure         \label{line:temporal-growth-end}
    \end{algorithmic}
    \end{scriptsize}
\end{algorithm}

\begin{algorithm}
    \caption{Bulk phase outer recursion function}\label{alg:bulk-outer-recursion}
    \begin{scriptsize}
    \begin{algorithmic}[1]
        \Statex
        \setcounter{ALG@line}{101}
        \Procedure{bulkRecursiveOuter}{$\delta, C$}
            \State $E, E^e, N^e, C^e, C^a \gets \emptyset, \emptyset, \emptyset, \emptyset, \emptyset$
            \For{$n \in \Call{sharedNeighbors}{C}$} \Comment{Process in increasing node label order} \label{line:outer-ext-start}  
                \State $\{u,v\} \gets C.X$
                \State $O_{uv} \gets$ \Call{outerOverlapClique}{$C, u, v, n, \delta$}
                \For{$(\{C_u, C_v\}, C_{new}) \in O_{uv}$}
                    \State $E \gets E \cup  C_{new}$   \label{line:outer-ext-end}
                    \If{$n > \max(C.X)$} \Comment{Limit branch cut logic to actual potential extensions}
                        \State $E^e \gets E^e \cup  C_{new}$ \label{line:outer-store-ext-start}
                        \State $N^e \gets N^e \cup n$
                        \State $C^e \gets C^e \cup \{(C_u,  C_{new}), (C_v,  C_{new})\}$ \label{line:outer-store-ext-end}
                    \EndIf
                \EndFor
                \State $C^a \gets C^a \cup \{C_k \in C^s |$ \parbox[t]{.6\linewidth}{ $C_k.X \in \{\{u,n\}, \{v,n\}\} \wedge$ \\ $C_k.\mathit{tbMin} \leq \max(\{C_i.\mathit{teMin}| (C_i, C_x) \in C^e\}) \wedge$ \\ $C_k.\mathit{teMax} \geq \min(\{C_i.\mathit{tbMax}| (C_i, C_x) \in C^e\})\}$} \label{line:outer-pot-select}
            \EndFor
            \Statex
            \State $D \gets \emptyset$ \label{line:outer-cut-start}
            \For{$(C_{n},  C_{new}) \in C^e$} \Comment{Determine which $C_{n} \in C^s$ can be cut}
                \If{\Call{isTemporallyDominated}{$C_{n},  C_{new}$}}  \label{line:outer-cut-newdom}
                    \State $E^r \gets \{C_k \in E^e | \max(C_k.X) \geq \max(C_{n}.X)\}$  
                    \If{\Call{isSpatialGrowthDominated}{$C_{n}, N^e, C_{new}, E^r$}}  \label{line:outer-cut-spatdom}
                        \State $N^r \gets \{u \in N^e| u \geq \max(C_{n}.X)\}$ \label{line:outer-temporal-growth-start}
                        \State $C^r \gets \{C_k \in C^a | \max(C_k.X) \geq \max(C_{n}.X)\}$ 
                        \State $C^p \gets \{C_i | (C_i, C_x)\in C^e \wedge \max(C_i.X) \leq \max(C_n.X)\}$ \label{line:outer-temporal-growth-mid}
                        \If{\Call{isTemporalGrowthDominated}{$C_{n}, E^r, N^r, C^r, C^p$}} 
                            \State $D \gets D \cup C_{n}$ \Comment{The branch defined by clique $C_{n}$ may be cut}
                        \EndIf \label{line:outer-temporal-growth-end}
                    \EndIf
                \EndIf
            \EndFor \label{line:outer-cut-end}
            \Statex 
            \State $\mathit{MAXIMAL}, D_{local} \gets True, \emptyset$
            \For{$C_{new} \in E$} \Comment{Extension cliques processed in increasing order based on $\max(X)$} 
                \If{\Call{isTemporallyDominated}{$C, C_{new}$}} \label{line:outer-check-maximal-start}  
                    \State $\mathit{MAXIMAL} \gets False$ 
                \EndIf  \label{line:outer-check-maximal-end}
                \If{$C_{new} \notin D_{local}$ and $\max(C_{new}.X) > \max(C.X)$}  \label{line:rec-conds-outer}
                    \State $D_{local} \gets D_{local} \cup \Call{bulkRecursiveInner}{\delta, C_{new}, E}$
                \EndIf \label{line:rec-conds-outer-end}
            \EndFor
            \Statex
            \If{$\mathit{MAXIMAL}$}  \label{line:outer-save-maximal-start}
                \State $R \gets R \cup C$
            \EndIf  \label{line:outer-save-maximal-end}
            \State \Return $D$
        \EndProcedure
    \end{algorithmic}
    \end{scriptsize}
\end{algorithm}

\begin{algorithm}
    \caption{Bulk phase outer recursion helper function, for determining node extensions from 2-node cliques with valid time interval overlap}
    \label{alg:bulk-outer-recursion-overlap}
    \begin{scriptsize}
    \begin{algorithmic}[1]
        \setcounter{ALG@line}{144}
        \Procedure{outerOverlapClique}{$C, u, v, n, \delta$}
            \State $O_{uv} \gets \emptyset$

            \For{$C_u \in \{C_k \in C^s | C_k.X = \{u, n\} \wedge$ \parbox[t]{.3\linewidth}{$C_k.\mathit{tbMin} \leq C.\mathit{teMin} \wedge$ \\ $C_k.\mathit{teMax} \geq C.\mathit{tbMax}\}$}} \label{line:outer-overlap-u-select}  
                \State $\mathit{tbMin^{new}} \gets \max(C.\mathit{tbMin}, C_u.\mathit{tbMin)}$ \Comment{Shrink potential growth}
                \State $\mathit{teMax^{new}} \gets \min(C.\mathit{teMax}, C_u.\mathit{teMax)}$ 
                \State $\mathit{tbMax^{new}} \gets \mathit{tbMin^{new}} + \delta - 1$
                \State $\mathit{teMin^{new}} \gets \mathit{teMax^{new}} - \delta + 1$
                \For{$C_v \in \{C_k \in C^s | C_k.X = \{v, n\} \wedge$ \parbox[t]{.3\linewidth}{$ C_k.\mathit{tbMin} \leq \mathit{teMin^{new}} \wedge$ \\ $C_k.\mathit{teMax} \geq \mathit{tbMax^{new}}\}$}} \label{line:outer-overlap-v-select}  
                    \State $\mathit{tbMin^{new}} \gets \max(\mathit{tbMin^{new}}, C_v.\mathit{tbMin)}$ \Comment{Shrink potential growth} \label{line:outer-overlap-vborder-select-start}
                    \State $\mathit{teMax^{new}} \gets \min(\mathit{teMax^{new}}, C_v.\mathit{teMax)}$ 
                    \State $\mathit{tbMax^{new}} \gets \mathit{tbMin^{new}} + \delta - 1$
                    \State $\mathit{teMin^{new}} \gets \mathit{teMax^{new}} - \delta + 1$ \label{line:outer-outborder-selected-end}
                    \Statex
                    \State $t_b^{uv}, t_b^{un}, t_b^{vn} \gets C.t_b, C_u.t_b, C_v.t_b$ \label{line:outer-overlap-interval-select-start}
                    \If{$t_b^{uv} < \mathit{tbMin^{new}}$} \Comment{Time interval $C$ outside outer border}
                        \State $t_b^{uv} \gets \min(\{t | (t,u,v) \in E^{\mathcal{T}} \wedge t \geq \mathit{tbMin^{new}}\})$
                    \EndIf
                    \If{$t_b^{un} < \mathit{tbMin^{new}}$} \Comment{Time interval $C_u$ outside outer border}
                        \State $t_b^{un} \gets \min(\{t | (t,u,n) \in E^{\mathcal{T}} \wedge t \geq \mathit{tbMin^{new}}\})$
                    \EndIf
                    \If{$t_b^{vn} < \mathit{tbMin^{new}}$}  \Comment{Time interval $C_v$ outside outer border}
                       \State $t_b^{vn} \gets \min(\{t | (t,n,v) \in E^{\mathcal{T}} \wedge t \geq \mathit{tbMin^{new}}\})$
                    \EndIf
                    \State $t_b^{new} \gets \min(t_b^{uv}, t_b^{un}, t_b^{vn})$ \Comment{New $t_b$ is first occurrence after $\mathit{tbMin^{new}}$}
                    \Statex
                    \State $t_e^{uv}, t_e^{un}, t_e^{vn} \gets C.t_b, C_u.t_b, C_v.t_b$
                    \If{$t_e^{uv} > \mathit{teMax^{new}}$} \Comment{Time interval $C$ outside outer border}
                        \State $t_e^{uv} \gets \max(\{t | (t,u,v) \in E^{\mathcal{T}} \wedge t \leq \mathit{teMax^{new}}\})$
                    \EndIf
                    \If{$t_e^{un} > \mathit{teMax^{new}}$} \Comment{Time interval $C_u$ outside outer border}
                        \State $t_e^{un} \gets \max(\{t | (t,u,n) \in E^{\mathcal{T}} \wedge t \leq \mathit{teMax^{new}}\})$
                    \EndIf
                    \If{$t_e^{vn} > \mathit{teMax^{new}}$}  \Comment{Time interval $C_v$ outside outer border}
                       \State $t_e^{vn} \gets \max(\{t | (t,n,v) \in E^{\mathcal{T}} \wedge t \leq \mathit{teMax^{new}}\})$
                    \EndIf
                    \State $t_e^{new} \gets \max(t_e^{uv}, t_e^{un}, t_e^{vn})$ \Comment{New $t_e$ is last occurrence before $\mathit{teMax^{new}}$} \label{line:outer-overlap-interval-select-end}
                    \Statex
                    \State $C_o \gets \mathit{Clique}($\parbox[t]{.6\linewidth}{$\{u,v,n\}, [t_b^{new}, t_e^{new}], $ \\ $(\mathit{tbMin^{new}}, \mathit{tbMax^{new}}, \mathit{teMin^{new}}, \mathit{teMax^{new}})))$} \label{line:outer-overlap-finish-start}
                    \State $O_{uv} \gets O_{uv} \cup (\{C_u, C_v\}, C_o)$
                \EndFor
            \EndFor
            \State \Return $O_{uv}$ \label{line:outer-overlap-finish-end}
        \EndProcedure
    \end{algorithmic}
    \end{scriptsize}
\end{algorithm}

\begin{algorithm}
    \caption{Bulk phase inner recursion function}\label{alg:bulk-inner-recursion}
    \begin{scriptsize}
    \begin{algorithmic}[1]
        \Statex
        \setcounter{ALG@line}{184}
        \Procedure{bulkRecursiveInner}{$\delta, C, E_{prev}$}
            \State $E, E^e, N^e, C^e, C^a \gets \emptyset, \emptyset, \emptyset, \emptyset, \emptyset$
            \For{$C_{p} \in E_{prev} \setminus \{C\}$}  \Comment{Process in increasing node label order} \label{line:inner-ext-start}  
                \State $O_e \gets$ \Call{innerOverlapClique}{$C, C_p, \delta$}
                \For{$C_{new} \in O_{e}$}
                    \State $E \gets E \cup C_{new}$ \label{line:inner-ext-end}
                    \If{$\max(C_p.X) > \max(C.X)$}
                        \State $E^e \gets E^e \cup C_{new}$  \label{line:inner-store-ext-start}
                        \State $N^e \gets N^e \cup \max(C_p.X)$
                        \State $C^e \gets C^e \cup (C_p, C_{new})$   \label{line:inner-store-ext-end}
                    \EndIf
                \EndFor
                \State $C^a \gets C^a \cup \{C_k \in E_{prev} |$ \parbox[t]{.6\linewidth}{ $\max(C_p.X) = \max(C_k.X) \wedge$ \\ $C_k.\mathit{tbMin} \leq \max(\{C_i.\mathit{teMin}| (C_i, C_x) \in C^e\}) \wedge$ \\ $C_k.\mathit{teMax} \geq \min(\{C_i.\mathit{tbMax}| (C_i, C_x) \in C^e\})\}$} \label{line:inner-pot-select}   
            \EndFor
            \Statex
            \State $D \gets \emptyset$ \label{line:inner-cut-start}
            \For{$(C_{n}, C_{new}) \in C^e$} \Comment{Determine which $C_{n} \in E_{prev}$ can be cut}
                \If{\Call{isTemporallyDominated}{$C_{n},  C_{new}$}}     \label{line:inner-cut-newdom}
                    \State $E^r \gets \{C_k \in E^e | \max(C_k.X) \geq \max(C_{n}.X)\}$ 
                    \If{\Call{isSpatialGrowthDominated}{$C_{n}, N^e, C_{new}, E^r$}}  \label{line:inner-cut-spatdom}       
                        \State $N^r \gets \{u \in N^e| u \geq \max(C_{n}.X)\}$              \label{line:inner-temporal-growth-start}
                        \State $C^r \gets \{C_k \in C^a | \max(C_k.X) \geq \max(C_{n}.X)\}$  
                        \State $C^p \gets \{C_i | (C_i, C_x)\in C^e \wedge \max(C_i.X) \leq \max(C_n.X)\}$ \label{line:inner-temporal-growth-mid}
                        \If{\Call{isTemporalGrowthDominated}{$C_{n}, E^r, N^r, C^r, C^p$}}
                            \State $D \gets D \cup C_{n}$ \Comment{The branch defined by clique $C_{n}$ may be cut}
                        \EndIf \label{line:inner-temporal-growth-end}
                    \EndIf
                \EndIf
            \EndFor \label{line:inner-cut-end}
            \Statex 
            \State $\mathit{MAXIMAL}, D_{local} \gets True, \emptyset$
            \For{$C_{new} \in E$} \Comment{Extension cliques processed in increasing order based on $\max(X)$}
                \If{\Call{isTemporallyDominated}{$C, C_{new}$}}  \label{line:inner-check-maximal-start} 
                    \State $\mathit{MAXIMAL} \gets False$ 
                \EndIf  \label{line:inner-check-maximal-end}
                \If{$C_{new} \notin D_{local}$ and $\max(C_{new}.X) > \max(C.X)$}    \label{line:rec-conds-inner}
                    \State $D_{local} \gets D_{local} \cup \Call{bulkRecursiveInner}{\delta, C_{new}, E}$
                \EndIf
            \EndFor
            \Statex
            \If{$\mathit{MAXIMAL}$}  \label{line:inner-save-maximal-start}
                \State $R \gets R \cup C$
            \EndIf  \label{line:inner-save-maximal-end}
            \State \Return $D$
        \EndProcedure
    \end{algorithmic}
    \end{scriptsize}
\end{algorithm}

\begin{algorithm}
    \caption{Bulk phase inner recursion helper function, for determining node extensions to $C.X$ with $\max(C_p.X)$ with valid time interval overlap}
    \label{alg:bulk-inner-recursion-overlap}
    \begin{scriptsize}
    \begin{algorithmic}[1]
        \setcounter{ALG@line}{226}
        \Procedure{innerOverlapClique}{$C, C_p, \delta$}
            \State $O_e \gets \emptyset$
            \State $X_{new} \gets C.X \cup C_p.X$
            \If{$C_p.\mathit{tbMin} \leq C.\mathit{teMin}$ and $C_p.\mathit{tbMax} \geq C.\mathit{tbMax}$} \label{line:inner-overlap-u-select}
                \State $\mathit{tbMin^{new}} \gets \max(C.\mathit{tbMin}, C_p.\mathit{tbMin)}$ \Comment{Shrink potential growth}
                \State $\mathit{teMax^{new}} \gets \min(C.\mathit{teMax}, C_p.\mathit{teMax)}$ 
                \State $\mathit{tbMax^{new}} \gets \mathit{tbMin^{new}} + \delta - 1$
                \State $\mathit{teMin^{new}} \gets \mathit{teMax^{new}} - \delta + 1$
                \Statex
                \State $a, b \gets \max(C.X), \max(C_p.X)$
                \For{$C_e \in \{C_k \in C^s | C_k.X = \{a, b\} \wedge$ \parbox[t]{.3\linewidth}{ $C_k.\mathit{tbMin} \leq \mathit{teMin^{new}} \wedge$ \\ $C_k.\mathit{teMax} \geq \mathit{tbMax^{new}}\}$}}  \label{line:inner-overlap-v-select}  
                    \State $\mathit{tbMin^{new}} \gets \max(\mathit{tbMin^{new}}, C_e.\mathit{tbMin)}$ \Comment{Shrink potential growth} \label{line:inner-overlap-vborder-select-start}
                    \State $\mathit{teMax^{new}} \gets \min(\mathit{teMax^{new}}, C_e.\mathit{teMax)}$ 
                    \State $\mathit{tbMax^{new}} \gets \mathit{tbMin^{new}} + \delta - 1$
                    \State $\mathit{teMin^{new}} \gets \mathit{teMax^{new}} - \delta + 1$ \label{line:inner-outborder-selected-end}
                    \Statex

                    \State $t_b^{uv}, t_b^{un}, t_b^{vn} \gets C.t_b, C_u.t_b, C_v.t_b$ \label{line:inner-overlap-interval-select-start}
                    \If{$t_b^{uv} < \mathit{tbMin^{new}}$} 
                        \State $t_b^{uv} \gets \min(\{t | (t,u,v) \in E^{\mathcal{T}} \wedge u,v \in C.X \wedge t \geq \mathit{tbMin^{new}}\})$
                    \EndIf
                    \If{$t_b^{un} < \mathit{tbMin^{new}}$} 
                        \State $t_b^{un} \gets \min(\{t | (t,u,v) \in E^{\mathcal{T}} \wedge u,v \in C_p.X \setminus C.X \wedge t \geq \mathit{tbMin^{new}}\})$
                    \EndIf
                    \If{$t_b^{vn} < \mathit{tbMin^{new}}$} \label{line:inner-tbvn-find-start}  
                       \State $t_b^{vn} \gets \min(\{t | (t,a,b) \in E^{\mathcal{T}} \wedge t \geq \mathit{tbMin^{new}}\})$
                    \EndIf \label{line:inner-tbvn-find-end}
                    \State $t_b^{new} \gets \min(t_b^{uv}, t_b^{un}, t_b^{vn})$ \Comment{New $t_b$ is first occurrence after $\mathit{tbMin^{new}}$}
                    \Statex
                    \State $t_e^{uv}, t_e^{un}, t_e^{vn} \gets C.t_b, C_u.t_b, C_v.t_b$
                    \If{$t_e^{uv} > \mathit{teMax^{new}}$} 
                        \State $t_e^{uv} \gets \max(\{t | (t,u,v) \in E^{\mathcal{T}} \wedge u,v \in C.X \wedge t \leq \mathit{teMax^{new}}\})$
                    \EndIf
                    \If{$t_e^{un} > \mathit{teMax^{new}}$} 
                        \State $t_e^{un} \gets \max(\{t | (t,u,v) \in E^{\mathcal{T}} \wedge u,v \in C_p.X \setminus C.X \wedge t \leq \mathit{teMax^{new}}\})$
                    \EndIf
                    \If{$t_e^{vn} > \mathit{teMax^{new}}$} \label{line:inner-tevn-find-start}  
                       \State $t_e^{vn} \gets \max(\{t | (t,a,b) \in E^{\mathcal{T}} \wedge t \leq \mathit{teMax^{new}}\})$
                    \EndIf \label{line:inner-tevn-find-end}
                    \State $t_e^{new} \gets \max(t_e^{uv}, t_e^{un}, t_e^{vn})$ \Comment{New $t_e$ is last occurrence before $\mathit{teMax^{new}}$}  \label{line:inner-overlap-interval-select-end}
                    \Statex
                    \State $C_{o} \gets \mathit{Clique}($\parbox[t]{.6\linewidth}{$X_{new}, [t_b^{new}, t_e^{new}],$ \\ $(\mathit{tbMin^{new}}, \mathit{tbMax^{new}}, \mathit{teMin^{new}}, \mathit{teMax^{new}})))$} \label{line:inner-overlap-finish-start}
                    \State $O_{e} \gets O_{e} \cup C_{o}$
                \EndFor
            \EndIf
            \State \Return $O_{e}$ \label{line:inner-overlap-finish-end}
        \EndProcedure
    \end{algorithmic}
    \end{scriptsize}
\end{algorithm}
\clearpage
\section{Supplementary result figures}
\vspace{-15pt}
\begin{figure}[H]
    \centering
    \begin{subfigure}[t]{\linewidth}
        \centering
        \includegraphics[width=\linewidth]{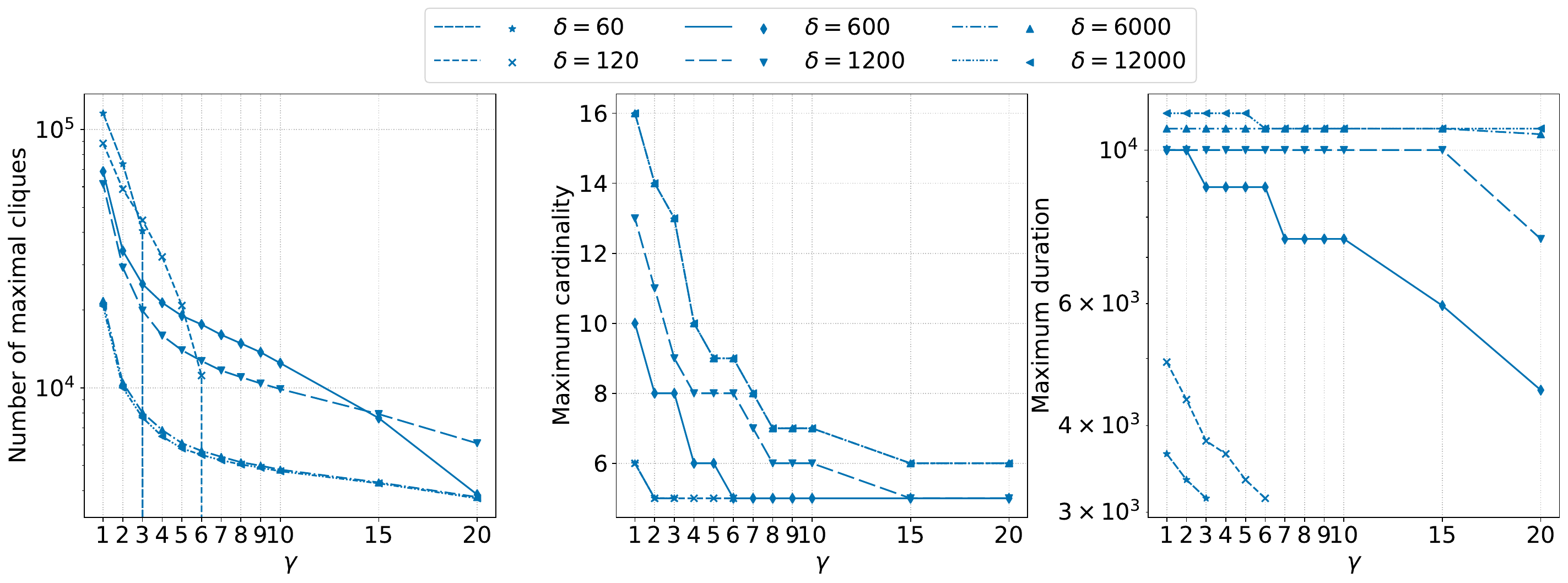}
        \caption{Infectious I}
    \end{subfigure}
    ~
    \begin{subfigure}[t]{\linewidth}
        \centering
        \includegraphics[width=\linewidth]{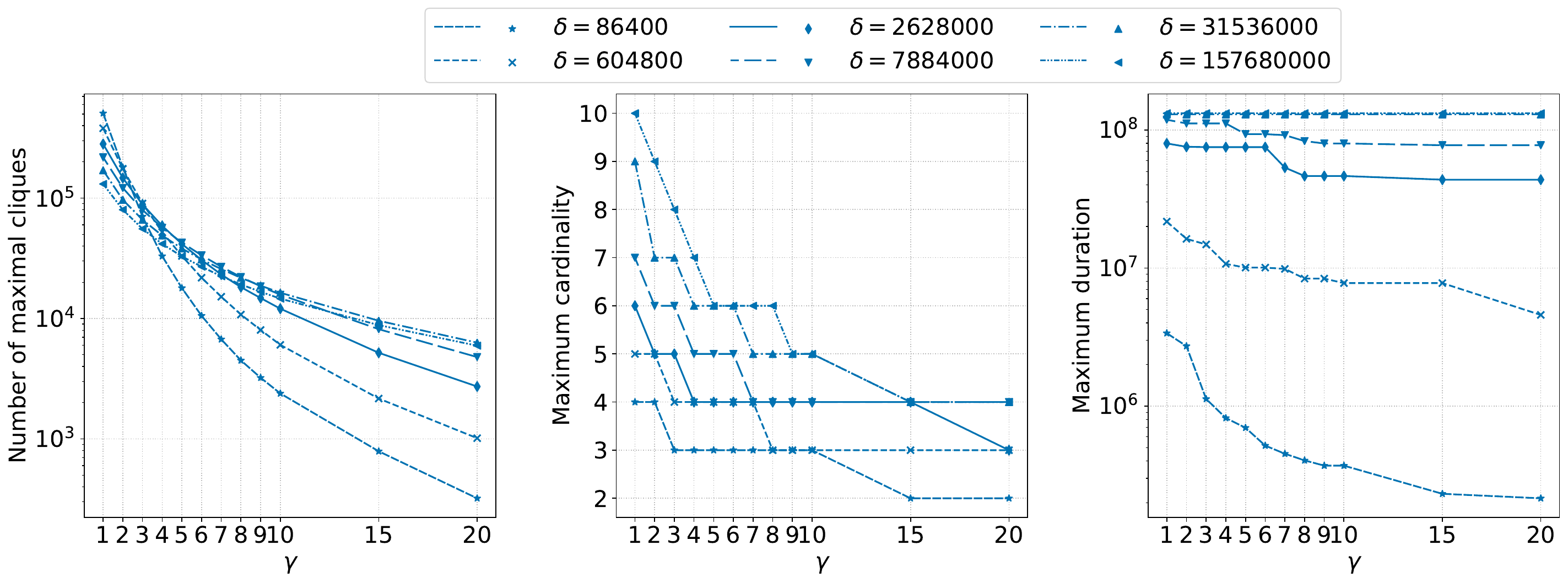}
        \caption{Facebook-wosn-wall}
    \end{subfigure}
    \begin{subfigure}[t]{\linewidth}
        \centering
        \includegraphics[width=\linewidth]{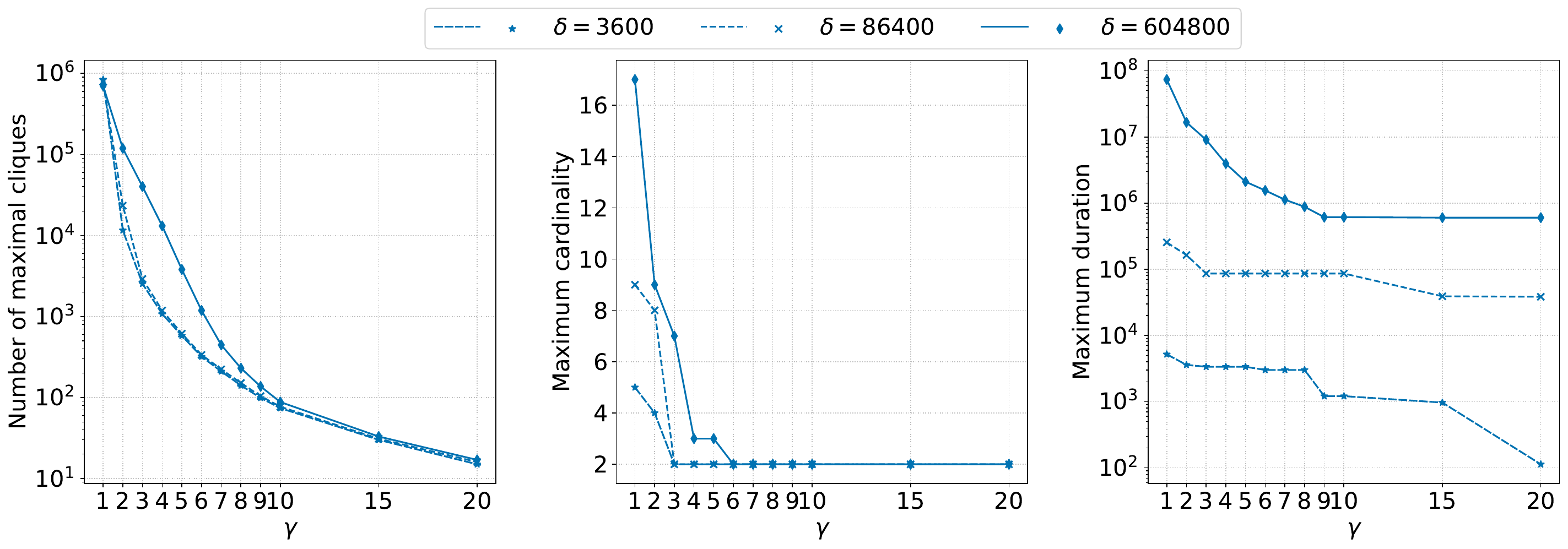}
        \caption{Reddit Hyperlinks}
    \end{subfigure}
    \caption{Properties of the set of enumerated cliques for a range of $\gamma$ values. The left panels show the number of ($\delta,\gamma$)-maximal cliques found and the middle and right panels show, respectively, their maximum cardinality and maximum duration.}
    \label{fig:res-gammacomp-properties-1}
\end{figure}

\begin{figure}[ht]
    \centering
    \begin{subfigure}[t]{\linewidth}
        \centering
        \includegraphics[width=\linewidth]{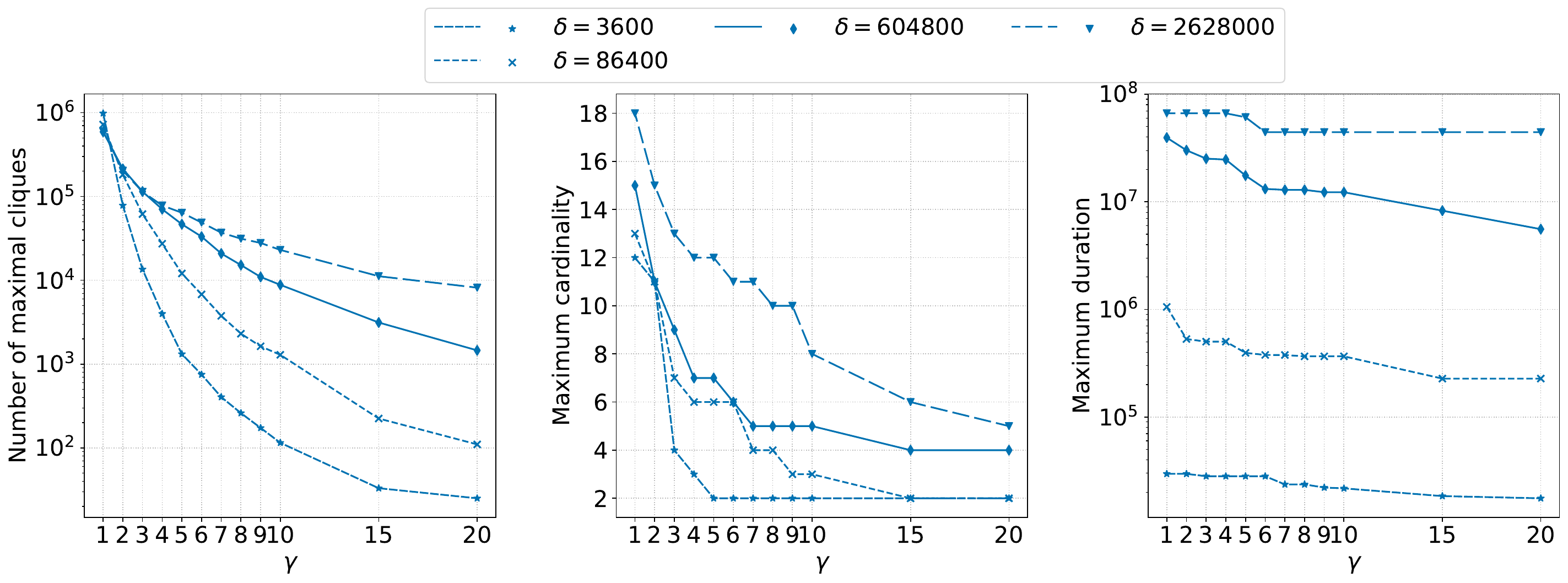}
        \caption{Enron}
    \end{subfigure}  
    \begin{subfigure}[t]{\linewidth}
        \centering
        \includegraphics[width=\linewidth]{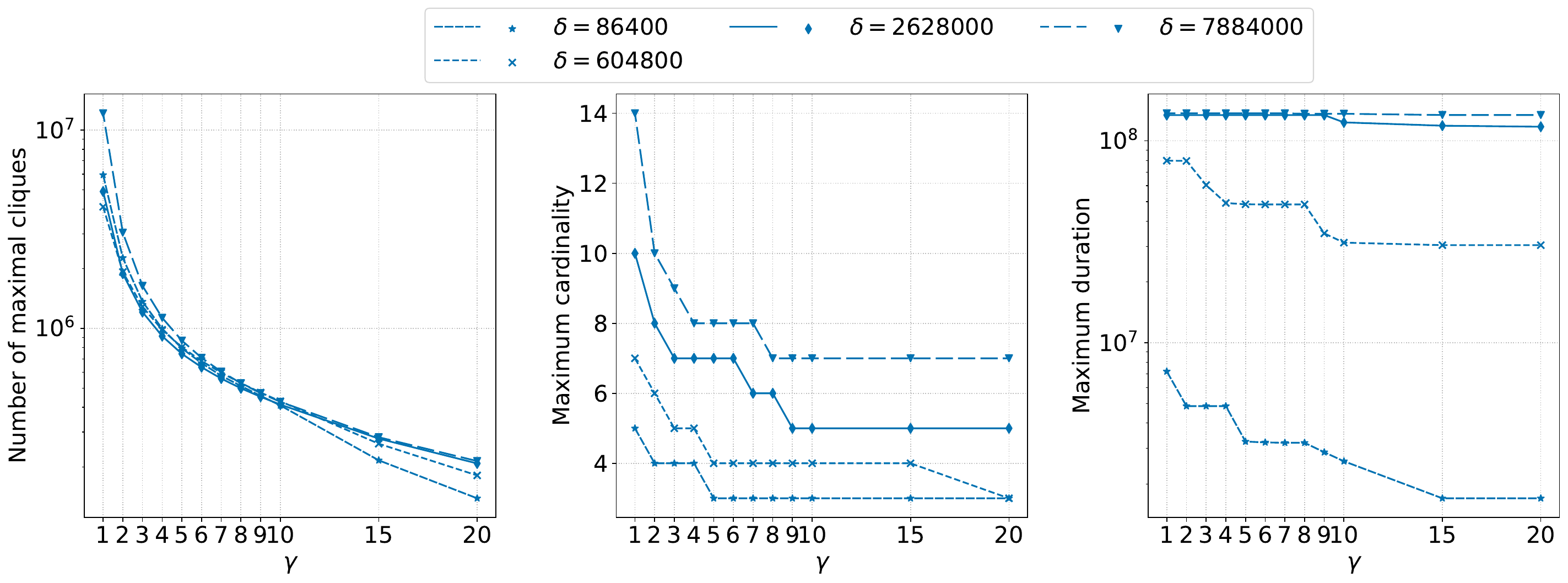}
        \caption{Last.fm bands}
    \end{subfigure} 
    ~
    \begin{subfigure}[t]{\linewidth}
        \centering
        \includegraphics[width=\linewidth]{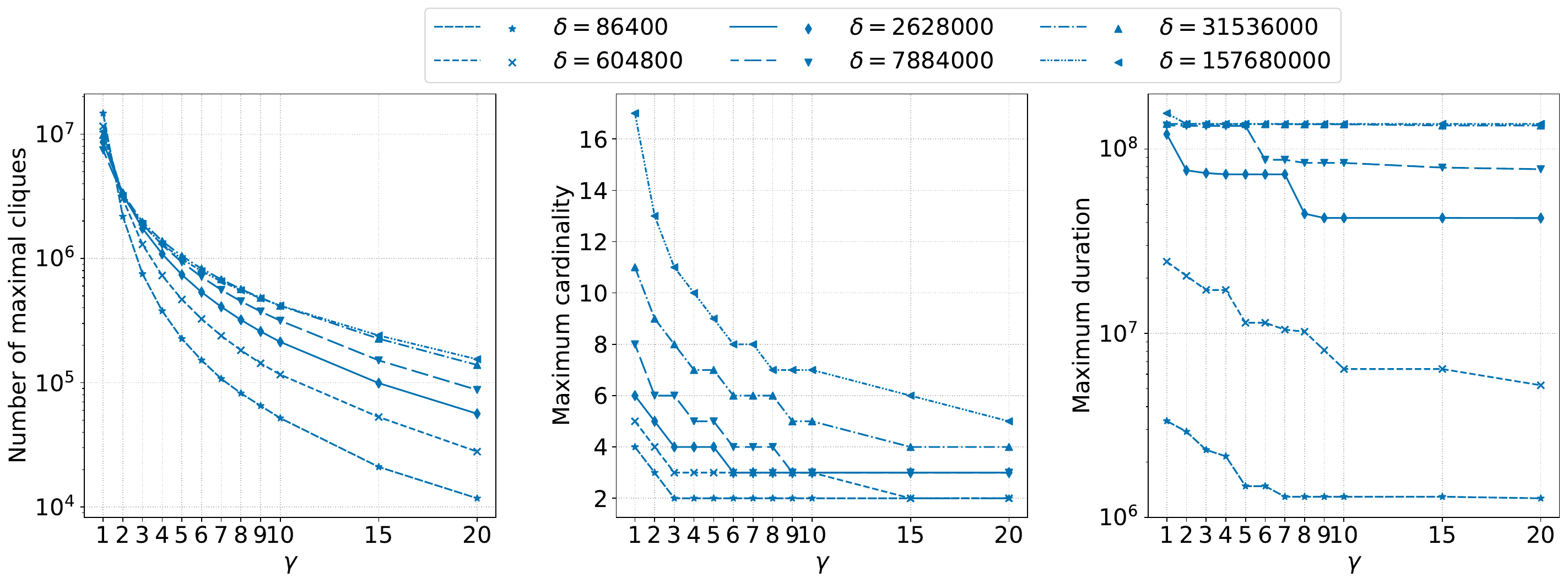}
        \caption{Last.fm songs}
    \end{subfigure}  
    \caption{Properties of the set of enumerated cliques for a range of $\gamma$ values. The left panels show the number of ($\delta,\gamma$)-maximal cliques found and the middle and right panels show, respectively, their maximum cardinality and maximum duration.}
    \label{fig:res-gammacomp-properties-2}
\end{figure}
\clearpage
\section{Experiments - Comparing the weighted and unweighted settings}
\label{apsect:results-weighted}
\begin{figure}[t]
    \centering
    \begin{subfigure}[t]{\linewidth}
        \centering
        \includegraphics[width=\linewidth]{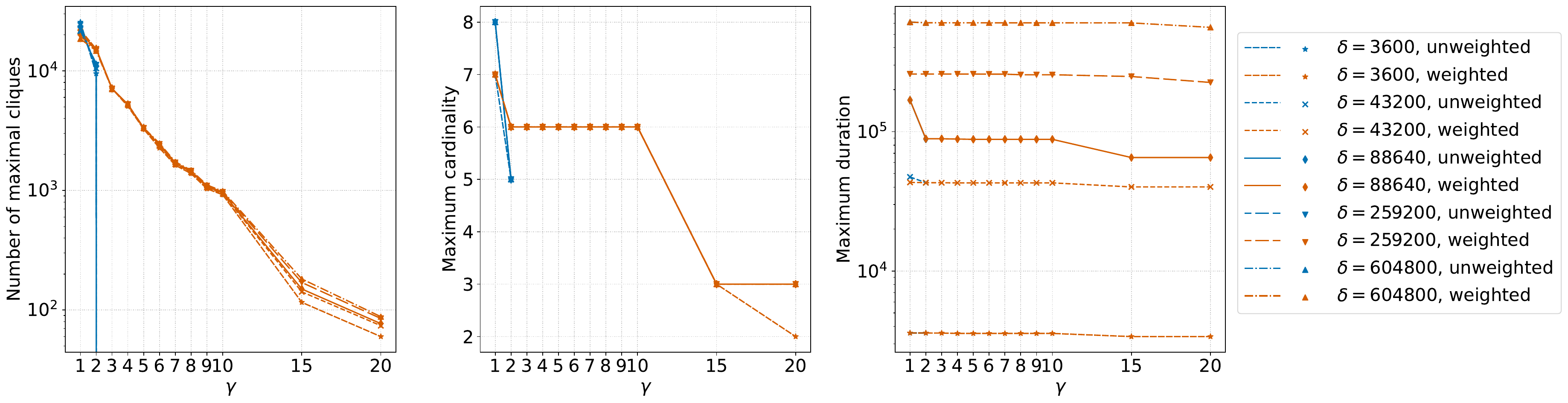}
        \caption{Bitcoin}
    \end{subfigure}
    \begin{subfigure}[t]{\linewidth}
        \centering
        \includegraphics[width=\linewidth]{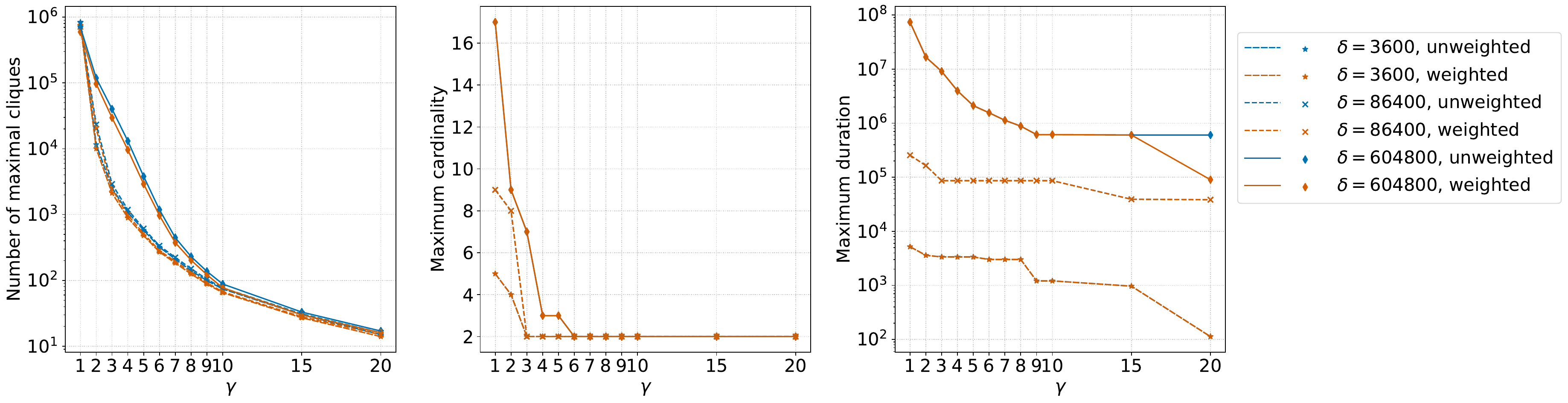}
        \caption{Reddit Hyperlinks}
    \end{subfigure}
    \caption{Properties of enumerated $(\delta,\gamma)$-maximal cliques, comparing the weighted and unweighted interpretation of two networks for the same $\delta$ and $\gamma$ values.}
    \label{fig:res-weighted-properties}
\end{figure}

To conclude our experiments, we compare the weighted and unweighted settings for the two weighted datasets (\texttt{Bitcoin} and \texttt{Reddit hyperlinks}), given identical $\delta$ and $\gamma$ values.
Figure~\ref{fig:res-weighted-properties} shows the number of maximal cliques, maximum cardinality, and maximum duration of the enumerated cliques.

For \texttt{Bitcoin} in the unweighted setting, we observe that there exist no node pairs with more than two edges connecting them within a week (our largest $\delta$).
As such, no more cliques are found at $\gamma > 2$.
On the contrary, given that \texttt{Bitcoin} weights range from -10 to 10, the weighted setting provides a larger range of results.
We observe that as $\gamma$ increases, thereby requiring increasingly positive relations between the clique members, we find fewer and fewer maximal cliques.
Thus, in the weighted setting, $\gamma$ can be used to fine-tune exactly how positively we would like the group members to rate one another, whereas in the unweighted setting we determine how frequently ratings must occur between members irrespective of their sentiment.

Unlike \texttt{Bitcoin} that had a range of weights, \texttt{Reddit hyperlinks} has but two possible weights, -1 or 1. 
As a consequence, the difference in results between the weighted and unweighted settings is much smaller. 
We observe that the weighted setting always finds fewer cliques, since cliques that include negative ties get excluded.
As such, whereas for \texttt{Bitcoin} the weighted and unweighted setting provided different uses, for \texttt{Reddit hyperlinks}, the weighted settings provides more of a refinement (or filtering) on the unweighted results.
Interestingly, in most cases this refinement does not seem to affect the maximum cardinality nor the maximum duration of the cliques.

\begin{figure}[ht]
    \centering
    \begin{subfigure}[t]{\linewidth}
        \centering
        \includegraphics[width=\linewidth]{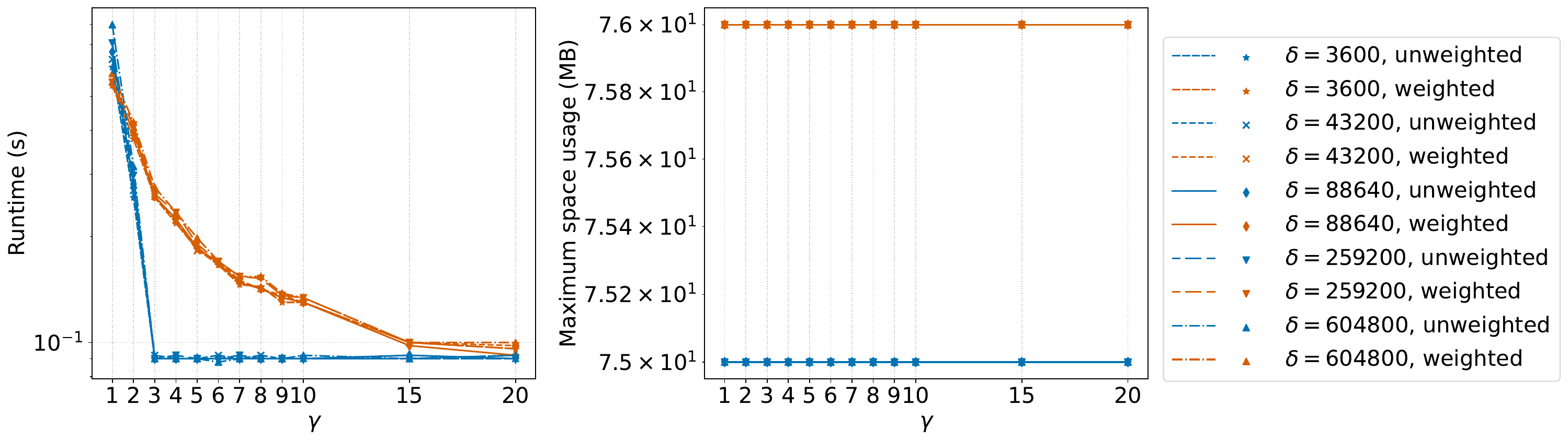}
        \caption{Bitcoin}
    \end{subfigure}
    \begin{subfigure}[t]{\linewidth}
        \centering
        \includegraphics[width=\linewidth]{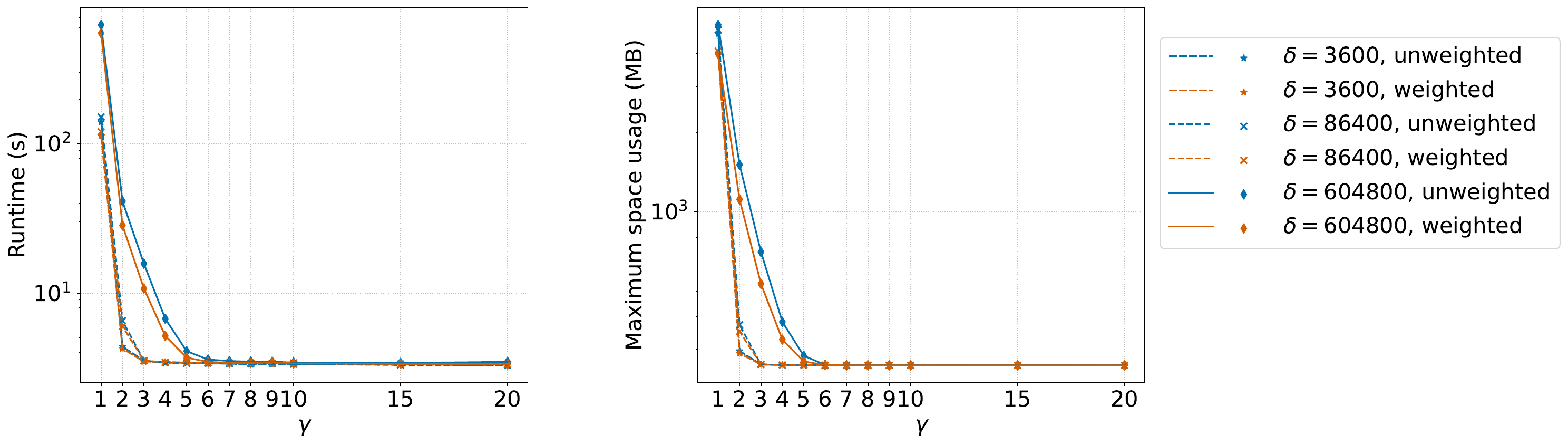}
        \caption{Reddit Hyperlinks}
    \end{subfigure}
    \caption{Performance of $(\delta,\gamma)$-maximal clique enumeration, comparing the weighted and unweighted interpretation of two networks for the same $\delta$ and $\gamma$ values.}
    \label{fig:res-weighted-performance}
\end{figure}

Performance results for the weighted and unweighted settings are shown in Figure~\ref{fig:res-weighted-performance}.
The figure shows that the slight decrease in $(\delta,\gamma)$-maximal cliques we observed for \texttt{Reddit hyperlinks}, corresponds to a similar slight decrease in runtime and space usage.
This confirms that $\gamma$ itself has no bearing on the runtime or space complexities of our algorithm.

\end{document}